\documentclass[useAMS,usegraphicx]{mn2e}

\newcommand{\be}{\begin{equation}}
\newcommand{\ee}{\end{equation}}
\newcommand{\bea}{\begin{eqnarray}}
\newcommand{\eea}{\end{eqnarray}}

\def\lesssim{\,\lower2truept\hbox{${<\atop\hbox{\raise4truept\hbox{$\sim$}}}$}\,}
\def\gtrsim{\,\lower2truept\hbox{${>\atop\hbox{\raise4truept\hbox{$\sim$}}}$}\,}

\title[Modelling SEDs with ANN]{Modelling the spectral energy distribution of galaxies: introducing the artificial neural network}

\author[L. Silva et al.]{
\parbox[t]{\textwidth}{
L. Silva$^1$\thanks{Email: silva@oats.inaf.it},
A. Schurer$^2$,
G.~L. Granato$^{1}$,
C. Almeida$^{3,4}$,
C.~M. Baugh$^{3}$,
C.~S. Frenk$^{3}$,
C.~G. Lacey$^{3}$,
L. Paoletti$^{5}$,
A. Petrella$^{5}$,
D. Selvestrel$^{5}$ }
\vspace*{6pt} \\
  $^1$INAF-OATs, Via Tiepolo 11, I-34131 Trieste, Italy \\
  $^2$School of GeoSciences, University of Edinburgh, Grant Institute, The King's Buildings, West Mains Road, Edinburgh EH9 3JW \\
  $^3$Department of Physics, Institute for Computational Cosmology, University of Durham, South Road, Durham DH1 3LE \\
  $^4$Key Laboratory for Research in Galaxies and Cosmology, Shanghai Astronomical Observatory, Chinese Academy of Sciences \\
  $^5$INAF-OAPd, Vicolo Osservatorio 5, I-35122 Padova, Italy \\}

\begin{document}
\date{Accepted ... Received ...}

\maketitle

\begin{abstract}
The spectral energy distribution of galaxies is a complex function of the star formation history and geometrical arrangement of stars and gas in galaxies. The computation of the radiative transfer of stellar radiation through the dust distribution is time-consuming. This aspect becomes unacceptable in particular when dealing with the predictions by semi-analytical galaxy formation models populating cosmological volumes, to be then compared with multi-wavelength surveys. Mainly for this aim, we have implemented an artificial neural network algorithm into the spectro-photometric and radiative transfer code \texttt{GRASIL} in order to compute the spectral energy distribution of galaxies in a short computing time. This allows to avoid the adoption of empirical templates that may have nothing to do with the mock galaxies output by models. The ANN has been implemented to compute the dust emission spectrum (the bottleneck of the computation), and separately for the star-forming molecular clouds and the diffuse dust (due to their different properties and dependencies). We have defined the input neurons effectively determining their emission, which means this implementation has a general applicability and is not linked to a particular galaxy formation model. We have trained the net for the disc and spherical geometries, and tested its performance to reproduce the SED of disc and starburst galaxies, as well as for a semi-analytical model for spheroidal galaxies. We have checked that for this model both the SEDs and the galaxy counts in the {\it Herschel} bands
obtained with the ANN approximation are almost superimposed to the same quantities obtained with the full \texttt{GRASIL}. We conclude that this method appears robust and advantageous, and will present the application to a more complex SAM in another paper.
\end{abstract}


\section{INTRODUCTION}
\label{section:intro}

The spectral energy distribution (SED) of a galaxy contains a wealth of information, and through its
study much can be learned about the galaxy's properties;
including the stellar and gas content of the galaxy, the age and
abundances of the stellar populations, the chemistry and physical state of
the interstellar medium, and the star formation rate (SFR) and history. It
is therefore the most direct probe to study galaxy formation and
evolution, both through direct observations and also by theoretical
modelling.

Different spectral ranges tend to be dominated by different specific
emission sources or radiative processes which affect the light as it
travels through the interstellar medium (ISM). Therefore by analyzing
and predicting the whole spectral range one can hope to deconvolve and
interpret all the different information contained in the SED in terms
of the SFR history and galaxy evolution in general. Stellar sources
mainly emit in the UV/optical to NIR spectral range, and the SED in
this wavelength region is therefore heavily influenced by the star
formation history of the galaxy and
as a result can be used to study the specific mixture of ages,
metallicity and mass distribution of the stellar populations. UV
photons ionize and excite the gas, producing HII regions with emission
lines, that are probes of the SFR and the chemistry, energetics and
physical state of the ISM where they are produced. Atomic and
molecular lines are present from the X-ray to the radio range
originating from electronic or rotational/vibrational transitions
(e.g. Stasinska 2007). The X-ray range probes mainly the
emission from hot plasma and from X-ray binary stars (e.g. Fabbiano
2006). The radio continuum emission is mainly produced by
free-free emission from ionized nebulae and synchrotron radiation by
energetic electrons accelerated in supernova remnant shocks and moving
in the galactic magnetic field (e.g. Condon 1992). The SED from a few
$\mu$m to the sub-mm (the IR region) is dominated by the interaction
of dust grains with stellar radiation. Dust in galaxies, although only
a small fraction of the mass of gas ($\sim 0.01$ in our Galaxy), is a
fundamental ingredient prevalent in many environments, such as
circumstellar envelopes, supernova remnants, star-forming regions and
diffuse clouds.  Dust grains absorb and scatter short wavelength
stellar radiation ($\lambda \lesssim 1 \mu$m) with high efficiency and
thermally emit the absorbed energy in the IR. In addition to its
effect on the SED, dust grains also affect many important chemical and
physical processes, for instance by acting as a catalyst for the
formation of H$_2$ molecules, by shielding dense and cold regions from
photo-dissociating UV photons allowing gravitational collapse and star
formation, by driving mass loss in evolved stars, and by depleting
heavy elements from the gas phase (e.g. Mathis 1990; Dorschner \&
Henning 1995; Draine 2003).

The modelling of the entire SEDs of galaxies is therefore very complex
and full of uncertainties. Because of this, several different
approaches have been proposed, depending also on the purpose of the
applications. Some works (e.g. Devriendt, Guiderdoni \& Sadat 1999;
Chary \& Elbaz 2001; Dale et al. 2001; Dale \& Helou 2002; Galliano et
al. 2003; Lagache, Dole \& Puget 2003; Da Cunha, Charlot \& Elbaz
2008) have proposed semi-empirical treatments of the SEDs. The aim of these works
is in general to interpret very large samples of data, requiring fast
computing times making use of observationally or physically motivated
SEDs. Other works are based on theoretical computations in
order to have a more general applicability in terms of interpretative
and predictive power. Within this approach different components and
levels of complexity have been considered. Several papers deal with
the radiative transfer (RT) in spherical geometries, mainly aimed at
modelling starburst galaxies (e.g.  Rowan-Robinson 1980;
Rowan-Robinson \& Crawford 1989; Efstathiou, Rowan-Robinson \&
Siebenmorgen 2000; Efstathiou \& Rowan-Robinson 2003; Takagi, Arimoto
\& Hanami 2003; Takagi, Vansevicius \& Arimoto 2003; Siebenmorgen \&
Krugel 2007). In the series of papers by Dopita et al.  (2005,
2006a,b) and Groves et al. (2008) a sophisticated modelling of the SED of
starburst galaxies have been presented, that includes the evolution of
stellar populations, the dynamical evolution of HII regions and
continuum and line emission. The series of papers by Popescu et
al. (2000), Misiriotis et al. (2001), Tuffs et al. (2004), Mollenhoff,
Popescu \& Tuffs (2006) are focussed on a detailed modelling and
intepretation of the SED of spiral galaxies, from the UV to sub-mm, to
provide constraints for several quantities such as optical depths,
attenuations, scale radii for the distribution of stars and dust. The
most general treatments of RT, capable of dealing with arbitrary
geometrical configurations, are based on Monte Carlo codes
(e.g. Bianchi et al. 1996; Gordon et al. 2001; Baes et al. 2003;
Chakrabarti et al. 2008; Li et al. 2008). Among these, the code
\texttt{DIRTY} (Gordon et al. 2001; Misselt et al. 2001) includes
extinction and dust emission and clumping of dust; \texttt{SUNRISE}
(Jonsson 2006; Jonsson \& Primack 2010; Jonsson, Groves \& Cox 2010)
computes extinction and dust thermal emission and has been applied to hydrodynamical
simulations of spirals including as sub-grid the HII region models by Groves et al. (2008);
the code \texttt{TRADING} (Bianchi 2007, 2008) includes both extinction and
dust thermal emission, the clumping of gas and stars, and has been
applied to study images of spirals. The drawback of Monte Carlo codes
is the very long computing times they require, which becomes
prohibitive when for instance applied to galaxy formation models in
cosmological volumes, where typically mock catalogues of many
thousands or tens of thousands of galaxies are necessary to compare
with observational constraints, for example multi-wavelength
luminosity functions (LF) and galaxy counts.

For a general purpose modelling of galaxy SEDs we developed the code
\texttt{GRASIL} (Silva et al. 1998 [S98]; Silva 1999 [S99];
Granato et al. 2000 [G00]; Bressan, Silva \& Granato 2002; Silva et al. 2001;
Panuzzo et al. 2003; Vega et al. 2005; Schurer et al. 2009). Our main aims were to construct a
relatively realistic and flexible multi-wavelength model, which could
calculate a galactic SED in a reasonably short computing time, to be
applied both to interpret observations and to make predictions in
conjunction with galaxy formation models. These requirements heavily
influenced our general choices, promoting the decision to include a
realistic bulge plus disk geometry, the radiative effects of different
dusty environments and the clumping of stars and dust, but to avoid Monte
Carlo calculations and to have some degree of geometrical (axial and
equatorial) symmetry. With these ingredients the model has been
successfully applied in many contexts (e.g. Granato et al. 2004; Baugh et al. 2005;
Silva et al. 2005; Panuzzo et al. 2007a,b; Iglesias-Paramo et al. 2007;
Fontanot et al. 2007, 2009; Galliano et al. 2008; Vega et al. 2008;
Lacey et al. 2008, 2010; Michalowski et al. 2009, 2010; Schurer et al. 2009;
Santini et al. 2010).

The study of galaxy formation and evolution has been receiving increasing
interest, both observationally and theoretically. Observational
programs covering the whole wavelength range are systematically and
directly unveiling galaxy populations at all redshifts, whose main
properties depend on the selection criteria. The detection of high
redshift galaxy populations are particularly important to track the
process of galaxy formation as a function of the cosmic epoch. Three main
spectral ranges are used, each detecting galaxies with different masses,
levels of star formation and at different evolutionary stages. IR and
sub-mm data collected initially by IRAS (Neugebauer et al. 1984; Soifer et
al. 1987), then mainly by COBE (Puget et al. 1996; Fixsen et al. 1998;
Hauser \& Dwek 2001), ISO (Kessler et al. 1996; Genzel
\& Cesarsky 2000), SCUBA (Holland et al. 1999; Smail et al. 1997, 2002;
Hughes et al. 1998) and more recently by {\it Spitzer} (Werner et al. 2004;
Soifer, Helou \& Werner 2008) have shown that at least half of the energy
emitted by stars over the history of the Universe has been reprocessed by
dust in the IR, with a high-z star formation activity much stronger than
locally, as witnessed by the fast evolution of the population of IR-bright
galaxies discovered with the mid-IR to mm cosmological surveys (see e.g.
the review by Lagache, Puget \& Dole 2005). At shorter wavelengths, large
populations of high-z star-forming galaxies, the Lyman-break galaxies
(LBGs), have been detected in the optical bands from their stellar
emission in the rest-frame UV, exploiting the spectral break around $912$
\AA\ produced by absorption by intervening neutral hydrogen (e.g. Giavalisco 2002).
The dust extinction corrections required
to provide an estimate for their SFR ($\sim$  $10$ to $100$ M$_\odot$/yr)
are probably very large (a factor $\sim 3-10$, e.g. Meurer, Heckman \&
Calzetti 1999), but remain very uncertain since at least part of the SFR
could be optically totally hidden. Deep near-IR surveys
and estimates of stellar mass functions
have revealed a substantial population of already massive, and in
many cases already evolved, galaxies at z $> 1$ (e.g. Drory et al. 2003, 2005;
Cimatti et al. 2004; Fontana et al. 2004; Bundy et al. 2005; Saracco et
al. 2005; Caputi et al. 2006). These observations reveal that the most massive galaxies tend to be
the oldest at all the sampled redshifts, i.e. the
high-luminosity/high-mass tails of the luminosity/mass functions are found
to evolve only weakly since $z \sim 5$ to now (e.g. Cimatti, Daddi \&
Renzini 2006).

From the theoretical point of view, the modelling of galaxy formation and
evolution in a cosmological context involves many processes at very
different scales, from Mpcs to a pc and under. The widest range of
observed galaxy properties have been analyzed using the so-called
semi-analytic models (SAM; White \& Rees 1978; Lacey \& Silk 1991; White
\& Frenk 1991), that consist in calculating the evolution of the baryon
component using simple analytical approximations, while the evolution of
the dark matter is calculated directly using gravity-only N-body
simulations, or Monte Carlo techniques based on the extended Press
Schechter theory (Lacey \& Cole 1993).

The final step to get the output simulated galaxy catalogues which can be
compared to observations, is the computation of the full wavelength range
SED for each mock galaxy. This should be calculated by appropriately
taking into account for each galaxy its particular star formation and
metallicity history and geometrical arrangement of the stellar populations
and of the ISM, as predicted by the model. The simulated SED catalogue can
then be compared to real observed galaxy surveys, so as to check whether
the predictions are or are not representative of the real universe and to
retrieve some information on the galaxy formation process. In principle,
the most general way to proceed would be to use a model which allows any
geometrical configuration for the distribution of stars and ISM, such as a
full Monte Carlo radiative transfer code. However, this is in practice not
feasible, because of unacceptable computing times, nor necessary since SAM
themselves lack detailed geometrical information about the simulated
galaxies. In fact, radiative transfer Monte Carlo codes at present are only
used coupled with hydro-simulations of single galaxies, not for
cosmological applications (e.g. Chakrabarti et al. 2008; Rocha et al. 2008; Narayanan et al. 2010).
Since the attempt to theoretically understand the
assembly of baryons within the hierarchy of dark matter halos is
inevitably subjected to strong uncertainties and degeneracies, as many
observational constraints as possible must be taken into account by
models, in order to get some hints as to the overall scenario and possibly
the main physical processes involved. Therefore only a complete
multi-wavelength analysis of galaxy data can be used to help unlock the
complexities of galaxy formation and evolution.

Most semi-analytical models have made use of simple empirical treatments
to compute the SED (e.g. Guiderdoni et al. 1998; Kauffmann et al. 1999;
Somerville \& Primack 1999; Hatton et al. 2003; Blaizot et al. 2004; Kang et al. 2005; Kitzbichler \& White 2007).
The only SAMs that includes a UV to
sub-mm radiative transfer computed from {\it first principle} are \texttt{GALFORM}
(Cole et al. 2000; G00; Baugh et al 2005; Lacey et al.
2008, 2010; Swinbank et al. 2008), \texttt{MORGANA} (MOdelling the Rise of GAlaxies aNd Active nuclei, Monaco
et al. 2007; Fontanot et al. 2007, 2009), and \texttt{ABC} (Anti-hierarchical
Baryonic Collapse; Granato et al. 2004; Silva et al. 2005; Lapi et al.
2006). These models have been interfaced with \texttt{GRASIL} to make detailed
comparisons and predictions in different spectral ranges.

As previously mentioned, the \texttt{GRASIL} code has been written in order to
calculate an accurate SED in a relatively quick time and this
has allowed the model to be used extensively for calculating the SEDs for
the above mentioned SAMs. Despite this, the calculation of the IR SED by
\texttt{GRASIL} is still often the bottleneck of the whole project and the
computing time becomes prohibitive when considering the exploitation of
large-scale structure simulations such as the Millennium Simulation
(Springel et al. 2005), which would require millions of galaxy
calculations.

To improve on this, with the idea in particular for use with cosmological
applications, we have implemented in \texttt{GRASIL} the possibility of computing
SEDs with an Artificial Neural Network (ANN) algorithm. This will reduce
the computing time significantly without having to rely on unrealistic
template approaches or simplified analytical recipes. According to the
required application one can choose the desired computational method: a
full \texttt{GRASIL} calculation or the quicker ANN mode. The bottleneck within the
\texttt{GRASIL} code is the computation of the cirrus and the molecular clouds dust
emissions. It is therefore these two processes that the ANN will be
applied to, with the option of using the ANN for either or both of the
processes. Another interesting application made possible due to the
improved performance of the ANN based computation would be the
combination of this code with an algorithm which could automatically
search the \texttt{GRASIL} parameter space in order to find optimized parameters to
fit real observed individual galaxy SEDs. In this first paper, the ANN for the emission from
the diffuse dust has been implemented for two geometrical arrangements, pure disc
or spherical distribution of stars and dust, and we test and apply it
to cases suited for these geometries. In particular, as a practical sample application, we compute galaxy counts in the
{\it Herschel} imaging bands for the \texttt{ABC} model for spheroidal galaxies. In an another paper, we will
present the implementation of the ANN also for the mixed bulge+disc geometry, and apply it to more complex
semi-analytical models (e.g. \texttt{GALFORM}).

Almeida et al. (2010) have already used the ANN algorithm to insert
the \texttt{GALFORM}+\texttt{GRASIL} model into the Millennium
Simulation, to study the properties of the population of sub-mm
galaxies. That method is complementary and substantially different
from the one presented here.  They identify the properties of the
\texttt{GALFORM} galaxies which determine, through an ANN, their
\texttt{GRASIL} SEDs. The method is successful and extremely fast,
since the ANN is used to compute the entire SED, not only the IR dust emission. However
their implementation is very specific to
\texttt{GALFORM}+\texttt{GRASIL}, and each realization of
\texttt{GALFORM} requires a training of the ANN, one for each output
redshift. The method we have implemented here is less fast but more
general, because the input is directly linked to the galactic
properties effectively determining the portion of the SED dominated by
dust emission (e.g optical depths, masses of dust, radiation field
etc.). As a result, one single training is able to cover a large
variety of applications.

In Sec.\ \ref{section:grasil} we recall the main properties of \texttt{GRASIL} and the
latest updates; in Sec.\ \ref{section:sedann} we
provide some generalities on ANN, 
and in Sec.\ \ref{section:anngrasil}
we describe the implementation of ANN in \texttt{GRASIL}, the choice of
the input neurons and the definition of the trained nets; in Sec.\ \ref{section:results}
we show some applications and examples. Finally our
conclusions are presented in Sec.\ \ref{section:concl}.

\section{Modelling SEDs with GRASIL}
\label{section:grasil}

\subsection{General description}
\label{section:gendes}

\texttt{GRASIL} (\texttt{GRA}phite $\&$ \texttt{SIL}icate) is a code constructed to
compute the SED of galaxies from the far-UV to the radio wavelength range,
treating with particular care the effects of dust reprocessing on the
stellar radiation and including continuum and line emission. In this
Section we provide a summary of its principal characteristics, that we
will need below to introduce the implementation of ANN. We refer to the
original papers for more details (in particular S98; S99; G00; Panuzzo et al. 2003; Vega et al. 2005).

The main aims of \texttt{GRASIL} are to provide a relatively realistic and
flexible modelling of galaxy SEDs, together with an acceptable (for most
applications) computing time. These requirements are reflected in its main
features:

\begin{itemize}
\item Galaxies are represented with stars and dust distributed in a
    bulge and/or a disc, adopting respectively a King and a double
    exponential profile (see e.g. Fig.\ 2.7 in S99 or Fig.\ 1 in G00  for a schematic
    representation of the geometry and components).
\item Three different dusty environments and their corresponding
    interaction with stars are considered: the star-forming molecular
    clouds (MCs) associated with newly-born stars, the diffuse medium
    (''cirrus'') associated with more evolved stars, and the dusty
    envelopes around AGB stars (intermediate age stars), their
    relative contribution to the SED depending on the star formation
    history.
\item The birth of stars within MCs and their gradual dispersion into
    the diffuse medium is accounted for by decreasing the fraction of
    energy stars emit within MCs with increasing age over a typical "escape
    timescale" (Sec.\ 2.5 and Eq.\ 8 in S98 for more details). Therefore we account for the
    clumping of (young) stars and dust within a diffuse medium, and for a
    greater attenuation suffered by the youngest stars, this means the
    attenuation is age-dependent (e.g. G00, in particular their Fig.\ 11,
    and Panuzzo et al.\ 2007a).
\item The dust model is made of graphite and silicate spherical grains with a continuous
    size distribution including grains in thermal equilibrium with the radiation field, very small grains
    fluctuating in temperature due to the absorption of single UV
    photons, and PAH molecules (optical properties by Draine \& Lee 1984; Laor \& Draine 1993;
    Li \& Draine 2001; Draine \& Li 2007). We compute the response to the incident
    radiation field for each type of grain.
\item The RT is exactly solved for the MCs with the Granato (1994,
    1997) code originally developed for AGN torii, implementing the
    $\Lambda$-iteration algorithm. These are represented as
    spherically symmetric clouds with the stars as a central point
    source (see Sec.\ 2.5.1 in S98 for a discussion on this assumption).
    Star forming MCs typically have extremely high optical depths even
    in the IR, which means IR-produced photons are self-absorbed, thus
    requiring a full RT treatment. Moreover, the youngest massive
    stars still embedded in MCs are also those emitting more strongly
    in the UV where the dust opacity is the highest.
\item The model galaxy is binned in appropriately small volume
    elements. The radiation field is evaluated in each of them from
    the knowledge of the distribution of stars and dust. Consequently
    the local dust emission and the attenuated radiation along each
    desired line of sight is computed. The treatment of the RT and
    dust emission in the diffuse phase (the real bottleneck of the
    whole computation) is approximated (see Section 2.5.2 in S98 and 2.5.3
    in S99).
\item Our reference library of SSPs is from Bressan et al. (1998,
    2002). We recall that the effects of the dusty envelopes around AGB stars and the radio emission (both thermal
    and non-thermal) are directly included in these SSPs. But any desired SSP library can be given in input to \texttt{GRASIL}
    (e.g. Fontanot \& Monaco 2010 tested the effects of both Bressan et al.\ and Maraston (2005) SSPs in \texttt{MORGANA}+\texttt{GRASIL}).
\item The output consists of a UV to radio SED. The maximum resolution is
    set by that of the input SSPs for the UV to NIR, while the wavelength
    points necessary to well define the dust features are in any case set
    by the code. In addition to the continuum and dust features, it is
    possible to include the computation of the nebular emission lines as described
    in Panuzzo et al. (2003).
\end{itemize}
It is worth noting that although the first release of \texttt{GRASIL} was more than
10 years ago, it is still the state of the art in the field. In fact the basic problem of RT
remains to find a compromise between computing time and choice of approximations, depending on the purpose.
We have carried out over the years several improvements, mostly following
observational progress. In particular, the emission bands from PAHs have
been updated with respect to S98, which were based on pre-ISO (Infrared
Space Observatory) observations, and Vega et al. (2005), based on ISO
observations (Li \& Draine 2001), by adopting at present the absorption cross sections and
band profiles by Draine \& Li (2007). This last update has been driven by
the availability of {\it Spitzer} data.

\subsection{Inputs}
\label{section:inpgra}

The inputs required by \texttt{GRASIL} consist of the star formation, gas and
metallicity evolution histories, and a set of geometrical parameters.

The former ingredients can be provided by analytical star formation laws, or by ``classical'' chemical
evolution models, or by more complex galaxy formation models (e.g.
\texttt{GALFORM}, \texttt{MORGANA}, \texttt{ABC}, see the Introduction).
Our reference code for generating star formation
histories, \texttt{CHE$\_$EVO}, is described in S99 (see also Section 2.2 in Fontanot et
al.\ 2009), and will be used below to generate the libraries to train the
ANN. It computes the evolution of the SFR, mass of gas, metallicity, and
of the chemical elements given an IMF and a SF law of the kind $SFR(t)= \nu \,
M_{gas}(t)^k + f(t)$, where the first term is a Schmidt-type SF with
efficiency $\nu$, and $f(t)$ is an analytical term.
Note however that our approach is independent of this choice, and
indeed our aim is to compute ANNs that can work with any engine to
generate the SF histories.

The other inputs required by \texttt{GRASIL}, some of which can be provided by galaxy formation models,
are:

\begin{itemize}
\item $f_{MC}$: mass gas fraction in star forming MCs. It affects mainly the FIR-submm.
\item $\tau_{MC}$: optical depth of MCs. It affects strongly the mid-IR and the silicate absorption feature.
\item t$_{esc}$: escape time scale of young stars from the MCs. It affects mainly the IR to UV-optical ratio.
\item $\delta \equiv$ M$_{dust}/$M$_{gas}$: dust to gas mass ratio. It is customary to set it either to a fixed value or proportional to the metallicity, unless provided by a dust evolution model (e.g. as in Schurer et al. 2009).
\item Bulge scale-radii (core radii of the King profile) and disk scale-radii and heights (of the double exponential) for stars and dust. The distribution of the radiation field relative to the dust determines the dust temperature distribution.
\end{itemize}

\section{Computing SEDs with Artificial Neural Networks}
\label{section:sedann}

\subsection{Basic concepts of ANN}
\label{section:basicann}

ANN were first introduced as very simplified models of the brain behavior
(McCulloch \& Pitts 1943; Rosenblatt 1958),
mathematical models able to learn from examples and data. They proved very
useful in tackling many computationally complex problems, generally
non-linear, such as pattern recognition, classification and function
approximation. They are now widely used in all scientific areas, for
instance in biochemistry, neuroscience, computer science, mathematics,
finance, physics as well as in astrophysics. The architecture of ANNs
reflects in someway the biological brain, in that they consists of
processing units (neurons) with multiple connections organized as a
network and working in parallel. These connections have adaptable
strengths (synaptic weights) which modify the signal transmitted to and
from each neuron. But in practice ANN can be considered powerful data
modelling tools with different possible implementations to address
different problems. Their ability to learn, generalization and
adaptability offer several advantages over other data mining and analysis
tools.

The working of ANNs is defined by their architecture, propagation rule and
learning algorithm:

\begin{itemize}

\item \textbf{Network Architecture}: The architecture or topology of
    ANNs  refers to the pattern of connections between the computing
    units and the propagation of  data. It can be split into 2 main
    classes, the {\it feed-forward} (FF, the kind of ANN we used) and
    the {\it feedback} or recurrent ANNs. In the FF case, the
    information moves only in the forward direction from the input to
    the output neurons. Recurrent networks contain feedback
    connections, cycles and loops. The neurons are commonly organized
    in {\it layers}, generally with an {\it input layer}, an {\it
    output layer} and in the more interesting case one or more {\it
    hidden layers}. In the FF pattern, each layer consists of units
    which receive their inputs from a layer directly below and send
    their output to units in a layer directly above. There are no
    connections within a layer. The simple network which can be built
    with no hidden layers is commonly called a {\it Perceptron}
    (Rosenblatt 1958), which can be used only for linear applications.
    For more difficult tasks, it is necessary to have at least one
    hidden layer (multi-layer perceptrons, henceforth MLPs). In
    particular, the {\it universal approximation theorem} (Haykin
    1999) states that one single layer of hidden units suffices to
    approximate any function to arbitrary precision, provided that the activation function (see the
    propagation rule) of the hidden layer is non-linear. Indeed, in
    our application we got satisfactory results with a single hidden
    layer.

\item \textbf{Signal propagation rule}: The basic working of the brain
    consists in neurons receiving electrochemical signals from other
    neurons, some of which excite the cell, while others inhibit it. The
    neurons add all these contributions and, if the sum is greater
    than a certain threshold, the neuron is activated, i.e. it
    transmits the signal further on. In analogy, each computational
    unit in the net receives a signal from all the neurons it is
    connected to, with the strength of each connection quantified by a
    weight. The unit multiplies each input signal by the weight of the
    corresponding connection and sums all the contributions. At least
    one non-linear {\it activation function} {\it f} is operated on
    the total signal to give the output value that is then passed on
    as an input to the neurons in the following layers. In practice:
    $o_j = f(\sum w_{jk} \, i_k)$ , where $o_j$ is the output signal
    from the $j$th neuron, $i_k$ are the incoming signals from all the
    neurons connected to it, with corresponding weights $w_{jk}$.
    Typical activation functions are the sigmoid,
    $f(x)=\frac{1}{1+e^{-c\,x}}$, Gaussian $f(x)=e^{-(c\,x)^2}$, and
    Elliot $f(x)=\frac{c\,x}{1+|c\,x|}$.

\item \textbf{Learning algorithms}: There are 2 main methods for the
    learning or training of the ANNs, {\it supervised} and {\it
    unsupervised} nets. In the first case (our case) the net is
    trained with a given target, i.e. the ANN is taught that for a
    given input it has to provide a given output, and the net adapts
    its connections (weights) so as to produce the desired answer. In
    unsupervised learning the net does not have a target output. It is
    used to find patterns and group the data.

    There are several methods within the supervised learning, all of
    which consist on a  comparison between the predicted output from
    the ANN with the target output. Our choice is the {\it
    back-propagation} (BP) algorithm (Rumelhart, Hinton \& Williams
    1986), which is the most widely used one. The errors are
    propagated backwards from the output nodes (directly defined by
    the comparison between the predicted and target values) to the
    inner nodes. This method is used to calculate the gradient of the
    error of the network with respect to the network's modifiable
    weights and thus to adjust the weights to find the (local) minimum
    of the error function with the gradient descent method.

\end{itemize}


{\bf In \texttt{GRASIL} we have implemented a standard feed-forward multi-layer back-propagation network (e.g. Bishop 1995; Rojas 1996).}

\subsection{Implementing ANN in GRASIL}
\label{section:anngrasil}

The prediction of galaxy spectra is a complex problem due to the high
number of input and output variables, and the non-linearity among them.
Neural networks represent a viable solution for this non-linear function
prediction.

The typical computing time to run \texttt{GRASIL} on a $\sim 2$ GHz CPU ranges from
$\sim$ half a second for the no dust case (i.e. the standard pure stellar spectral synthesis which is not our interest here) to $\sim 10$ minutes for the bulge plus disc case. A pure bulge (i.e. spherical symmetry) requires
$\gtrsim 1$ minute, a pure disc $\gtrsim 2$ minutes, a combined bulge and disc
geometry requires $\sim 10$ minutes (the exact value depending on the
number of radial and angular grid points set according to the
''compactness'' of the model, see S98 and S99 for details). Most time is
spent computing the IR emission from dust, of which $\sim 70$ to $> 90$\%
is required by the emission from the cirrus component. This is because each
volume element of the model in a 3D grid has its own radiation field and
amount of dust, whose emission is calculated individually for each type of grain of the dust size distribution,
including grains in thermal equilibrium with the radiation field, and small grains and PAH requiring the computation of a temperature probability distribution function.
Instead, computing the extinction of stellar radiation by the two dusty
components of the ISM (molecular clouds and cirrus) is straightforward,
$\sim 1$ second.

These considerations drove our strategy to use an ANN
algorithm to reconstruct only the IR emission from molecular clouds and
cirrus. To this aim, the fast calculation of the extinction by the
molecular clouds and the cirrus provides the amount of energy absorbed and
therefore the normalization for the two components, while we have
implemented the ANN to predict their spectral shapes. Since the CPU time
required to predict the IR emission with the ANN is negligible with
respect to the direct computation, with this approach we can expect orders
of magnitude gain in the time performance.

Due to the very different nature and treatment of the MCs and cirrus
components, the quantities (input neurons) that determine their respective
IR SEDs are different. Therefore we have implemented a distinct
ANN for each of them. As a byproduct, it is possible to run the code in
the ``ANN mode'' for both components or for only one of them. In the
latter case, the emission from the other component is computed in the
``full'' mode.

We have implemented a {\bf standard feed-forward back-propagation} MLP with one hidden layer, using a sigmoid
activation function from the input to the hidden layer. Indeed by properly
identifying the input neurons and setting the network parameters (number
of neurons of the hidden layer, learning rate, maximum number of
iterations) the universal approximation theorem implies that this
architecture is capable of universal optimal approximation. Moreover, it
is found empirically that networks  with multiple hidden layers are more
prone to getting caught in undesirable local minima (Haykin 1999; Bishop 1995).
To create and use the trained net to predict
SEDs we have adapted the F90 code by B. Fiedler freely available at
{\it http://mensch.org/neural/}.

\subsection{Input and output neurons}
\label{section:neurons}

For each dusty component we have identified the physical quantities
controlling their spectral shape. Their integrated luminosity, i.e.\ the
normalization, is known from the direct computation of the amount of
stellar energy absorbed by MCs and cirrus respectively.

The identification of the input neurons is based on physical expectations
corroborated by working experience with \texttt{GRASIL}. As such, they are closely
related to, but not coincident with, \texttt{GRASIL} parameters. This is because
different combinations of two or more parameters produce identical
or very similar dust emission SEDs. For instance, different combinations
of dust to gas ratio $\delta$, MC mass $M_{MC}$ and radius $R_{MC}$
produce the {\it same} MC SED as long as the MC optical depth
$\tau \propto \delta M_{MC}/R_{MC}^2$ is unchanged (S98, S99). The same is true
for the shape of the cirrus SED, if all the relevant masses (cirrus dust
mass and SFR) are varied by a factor $f$ and, at the same time, the scale
radii by $\sqrt f$. Also, the cirrus, and even more, the MC  dust
emission depend only weakly on details of the spectral shape of the input
stellar radiation, which means that different combinations of SFR$(t)$,
$Z(t)$, galactic age $T_{gal}$, and MC escape timescale $t_{esc}$ (which
affects the fraction of starlight heating the MC and that heating the
cirrus) may well give rise to almost identical dust emission in one or
both components.

The output neurons are the values of $\lambda L_\lambda$ in the IR
region, both for molecular clouds and for the cirrus, which usually
means several hundred output neurons.

\subsubsection*{Input neurons for molecular clouds}

Since in all practical cases the MCs are optically thick to the primary
stellar emission heating them, and they are approximated by homogeneous
spheres with constant density, the shape of the emitted IR SED is
controlled only by the two quantities listed below, which are simply
related to \texttt{GRASIL} parameters. Accordingly, the adopted input neurons are:

\begin{itemize}
\item $\tau_{MC} \propto \delta \,M_{MC}/R_{MC}^2$, the  molecular cloud
    optical depth (conventionally given at 1 $\mu$m).
\item $R_{MC}[pc]/(k \sqrt(L_{\star MC,46})$ the ratio of the
    molecular cloud radius over an estimate of the dust sublimation radius,
    i.e.\ the inner radius of the dust distribution. The latter
    depends on the luminosity of stars within each cloud (in
    $10^{46}$ erg/s). The constant $k$ depends on the adopted maximum
    temperature $T_s$ for dust. We found (see discussion in S98 and
    S99) that a value of $T_s=$ 400 K, properly represents the maximum attainable
    dust temperature in star forming molecular clouds, for which
    $k\simeq 12$. We explicitly note that the value of this constant
    is irrelevant as long as we use the same value when training and
    using the MLP. In other words, any quantity proportional to this
    ratio would be equally good as input neuron.
\end{itemize}

\begin{figure}
  \centerline{
  \includegraphics[width=10cm]{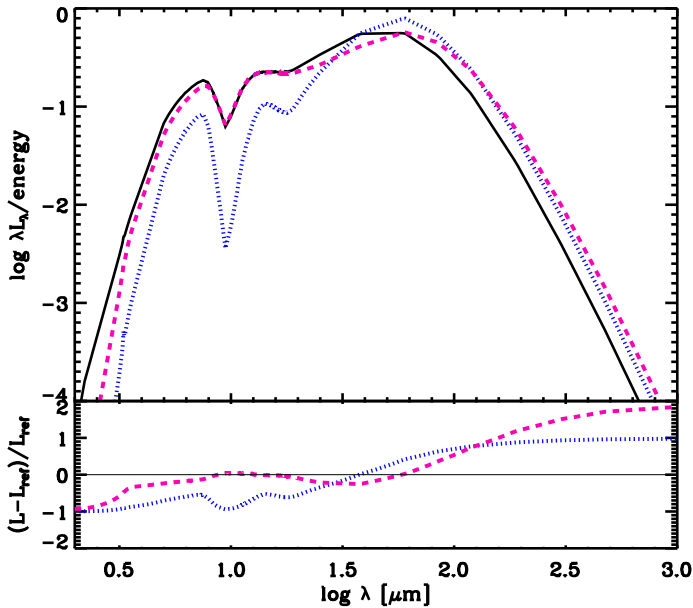}
    }
  \vspace{-3.4cm}
\caption{Example of the effects of the controlling parameters of the
MC SED. Upper panel. Continuous (black) line: reference SED ($\tau_{MC} = 25.4$
at $1 \mu$m, and $R_{MC}[pc]/\left (12 \sqrt{L_{\star MC,46}}\right )
= 104$). Dotted (blue) line: effect of varying $\tau$ alone by a
factor of 2 (obtained by doubling the dust to gas mass ratio). Dashed
(pink) line: effect of varying the molecular cloud extension alone by
a factor of 2 (obtained by decreasing the escape time scale, i.e.\ the
primary source luminosity within MCs and therefore the inner radius,
to a value able to provide a factor of 2 increase of the neuron). The
SEDs are normalized to their own energy to highlight the change in
shape. Lower panel. Residuals with respect to the reference SED.} \label{fig:mcinput}
\end{figure}

In Fig. \ref{fig:mcinput} we highlight the separate effect of the two
input neurons in the SEDs of molecular clouds. Their variation is
obtained by changing values of different parameters, as detailed in
the caption. The optical depth has a strong effect in the slope of the
MIR and on the depth of the silicate absorption. In the example depicted,
with the starting value of $\sim 25$ at $1 \mu$m, the IR emission is
self absorbed up to $\sim 25 \mu$m. By doubling it, it becomes $<$ 1
at $\sim 40 \mu$m, with a much stronger self absorption giving rise to
a steeper slope. The increase in the extension alone allows colder
dust temperatures in the outskirts of the MCs and therefore emission
at longer wavelengths.

\subsubsection*{Input neurons for Cirrus}

The cirrus emission is defined by 6 neurons for spherical symmetry and 9
for discs, listed below. As for mixed bulge and disc geometry, we have found
that if dust heating is dominated by stars in the
disc component, as is the case for nearby spirals, the pure disc network gives sufficiently
accurate results. But for a general application to mock galaxies output by SAMs,
a mixed geometry must be available, unless the model explicitly takes into account only spheroidal or disc galaxies.
As recalled in the Introduction, in this paper we present the implementation of the ANN for spherical or disc geometries,
while the application to the mixed geometry will be presented in another paper (in preparation).

\begin{itemize}

\item $\log (L_{Cir}/L_{\star,c})$, the cirrus dust luminosity
    normalized to the stellar luminosity heating the cirrus. The
    former can obviously be derived without actually computing the
    dust emission, since it equals the stellar energy absorbed by the
    cirrus. This ratio provides a global measure of the amount of dust
    reprocessing.

\item $\log (M_{Cir}/L_{\star,c})$, the normalized cirrus dust mass,
    expected to be strongly correlated with the (distribution of) dust
    emitting temperature.

\item $\tau_p$ and $\tau_e$, the polar and equatorial optical depths
    due to cirrus alone (integral of the dust density distribution
    along the polar and equatorial directions respectively,
    conventionally given at $1 \mu$m). Only one of the two is used for
    pure bulge geometry, since $\tau_e=\tau_p$.

\item $\tau_h$, a fictitious optical depth, computed as if cirrus were
    spherically and homogeneously distributed. This dummy quantity was
    already computed by \texttt{GRASIL}, and it is included here because its
    comparison with $\tau_p$ and $\tau_e$ provides a measure of the ``concentration''
    of the dust distribution independently of the specific density law
    assumed. Of course this concentration significantly affects the
    shape of the emitted SED, and indeed we empirically found that its inclusion
    improves the performances of the MLP.

\item \textbf{Geometrical ratios}: Dependent on the geometry of the
    galaxy used. For the bulge component:
    $r_{c,*}/r_{c,diff}$ which measures the ``relative position'' of dust
    and stars. For the disc component $r_{d,*}/r_{d,diff}$,
    $z_{d,*}/r_{d,*}$, $z_{d,diff}/r_{d,diff}$. Taken together these
    three ratios measure the relative position of dust and stars and
    the geometrical thickness of star and dust distributions.

\item \textbf{Hardness ratio}: ratio of the radiation field at 0.3
    $\mu$m over 1 $\mu$m, heating the cirrus (thus emerging from
    molecular clouds and stars already out of molecular clouds). Since
    small grains and especially PAHs are excited most effectively by
    relatively hard UV photons, this quantity is correlated with the
    ratio between the NIR-MIR emission they produce, and the far-IR
    due to big grains.

\end{itemize}

\begin{figure}
  \centerline{
   \includegraphics[width=10cm]{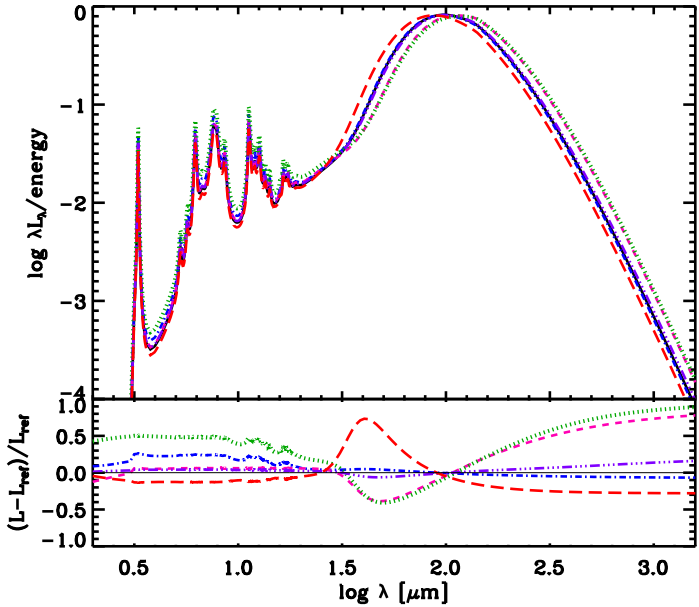}
    }
\vspace{-3.4cm}
\caption{Examples of the effect of the cirrus input neurons. Continuous (black) line: reference cirrus SED for a spherical model.
Dot-dashed (blue) line: hardness ratio increased by $\sim 20$\% (obtained by avoiding the SSPs with the highest metallicy available in our library, so to have somewhat harder stellar intrinsic spectra). Short dashed (pink) line: increase M$_{dust}$ by a factor of
2 (obtained by doubling $\delta$, and the scale radii by $\sqrt(2)$ to leave $\tau$ unchanged). Dotted (green) line: decrease $\tau$ of
cirrus dust by a factor 5 (by increasing stars and dust scale radii by the same factor to leave their ratio and the dust mass unchanged).
Three-dot-dashed (violet) line: decrease $\tau_h$ by a factor 10 (by increasing the galaxy radius by $\sqrt(10)$ to leave $\tau$ essentially unchanged since the King profile is quite centrally concentrated). Long-dashed (red) line: star to dust scale radii ratio halved (by halving the stellar scale radius).}
\label{fig:cirinput}
\end{figure}

Examples of the effect of the cirrus neurons are shown in Fig.\ \ref{fig:cirinput}. A variation in the hardness ratio, also by only $\sim 20$\% as in the figure, has an immediate effect in the temperature distribution of dust grains with small heat capacity, i.e. very small grains and PAHs, therefore it mainly affects the mid-IR emission leaving almost unchanged the far-IR. Increasing the amount of dust alone has greater effects in the overall equilibrium temperature of dust grains, and therefore in the position of the peak of the FIR emission, because of a smaller photon to dust density ratio. A similar effect in the FIR coupled to a hotter MIR as in the first case, can be obtained by lowering the optical depth of the dust distribution leaving unchanged the amount of dust. Indeed in this case, on one side a lower concentration of the radiation field relative to the dust density yields colder equilibrium temperatures, on the other side small grains and PAHs respond to single UV photons, which have a longer mean free path with a smaller $\tau$. The effect of $\tau_h$ alone is a modulation with respect to $\tau$. The shape of the SED has a strong dependence on the star to dust scale radii, since this implies a different distribution of the radiation field and therefore a redefinition of the temperature distribution function of dust grains. We note that the input neurons we found to work for the ANN are not fully independent, in fact a variation of essentially anyone of them implies also different amounts of reprocessing. In the depicted examples, the reprocessing changes from a few percents to $\sim 40$\%. We empirically found that the ANN provides a better performance with this additional information.

\subsection{Network training}
\label{section:train}

The MLPs we use in the following have been trained on some thousands
\texttt{CHE$\_$EVO}+\texttt{GRASIL} models\footnote{The training libraries have been
efficiently produced using \texttt{GRASIL} WEB interface GALSYNTH, accessible
through http://adlibitum.oat.ts.astro.it/silva/default.html}, either for pure
spheroids or pure discs, covering generously the range of parameter values
used in several of our past works. Actually, the definition of how large the range must be is non trivial,
since the properties of the mock galaxies calculated in simulations of
galaxy formation are not predictable a priori, nor are those of high-z
galaxies in the real universe.

{\bf For the applications shown in the next Section, the range of values of the input neurons used for the training is very large, this is particularly required for the ABC galaxy model, characterized by the presence of extreme phases of the evolution of the SFR (Sec.\ \ref{section:abc}).
Specifically:
\begin{itemize}
\item $\tau_{MC}$ = $1$ to $70$;
\item $R_{MC}/R_{min}$ = $18$ to $3000$;
\item $\log (L_{Cir}/L_{\star,c})$ = -2 to -0.02;
\item $\log (M_{Cir}/L_{\star,c})$ = -8 to -6;
\item $\tau_e$ = 0.01 to 100;
\item $\tau_p$ = 0.007 to 70;
\item $\tau_h$ = 8e-4 to 8;
\item $r_{c,*}/r_{c,diff}$ and $r_{d,*}/r_{d,diff}$ = 0.2 to 5;
\item $z_{d,*}/r_{d,*}$ and $z_{d,diff}/r_{d,diff}$ = 0.02 to 0.3;
\item $\log (L_{\star,c}(0.3)/L_{\star,c}(1))$ = -0.6 to 0.8.
\end{itemize}
}

\begin{figure}
  \centerline{
   \includegraphics[width=10cm]{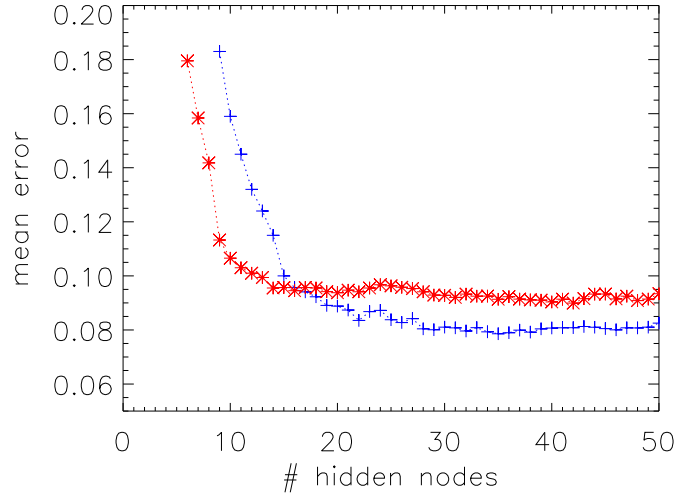}
    }
\caption{{\bf Mean total error (i.e. averaged over all wavelengths and and all models used for the training) provided by the trained network on the normalized (-1 to 1) fluxes of the verification sets, as a function of the number of neurons in the hidden layer, for molecular clouds (asterisks) and cirrus (crosses).}}
\label{fig:nnodi}
\end{figure}

{\bf We trained the net with 90\% of the models, randomly chosen within the library generated with the aforementioned range of parameters, and using the remaining 10\% as a verification set. The training procedure does not (in this application) change the structure of the NN.} We have empirically adjusted the number of neurons in the hidden layer $n_{hid}$. {\bf As shown in Fig.\ \ref{fig:nnodi}, the error provided by the network on the verification sets decreases with increasing number of nodes down to a minimum, then it remains substantially flat unless the number becomes very large. In other words, the increase after the minimum is very shallow if any, and then the minimum point is not very well defined. Guided by this and similar plots on different test set selections, in the applications shown below we have adopted $n_{hid}=20$ for molecular clouds and 35 for cirrus}. We trained the MLP using 500 training epochs (iterations) with a learning rate of 0.001. Also these choices are not very critical for the final results.

An important expected advantage of the ANN technique with respect to
classical interpolations is the capability to ``learn'' the effect of each
single input neuron on the SED, mimicking in some sense the skill that a
real \texttt{GRASIL} user develops with experience. Therefore we may expect that
the MLP can produce a correct SED, corresponding to a given choice of
input neurons, even when the trained set does not include examples with the
{\it entire} set of inputs neurons bracketing the required ones at the
same time. It is normally sufficient that each single input neuron is
independently within the values included for the training. Indeed, the
general performance we have experienced confirms our expectations.
On the other hand, we expect and we found that the ANN often fail catastrophically, when one or
more parameters are not within the range of values of the training set.
{\bf For a given input galaxy model, it is therefore fundamental to check if the training encloses all the values of the input neurons. Actually, to prevent the general user from an improper use of the ANN, we have implemented in \texttt{GRASIL} a check to force the full radiative transfer computation whenever the previous condition is not met.}


Note that the ANN adopted in the following has been trained on
models computed with given intrinsic dust properties, {\bf specifically properties compliant with the average Milky Way-type dust} (size distributions, relative abundances of graphite and silicates, PAH abundance, slope of the dust emissivity in the sub-mm etc, see S98).
The full \texttt{GRASIL} model has the freedom to modify these
quantities, even if in the standard use this is not usually
exploited, {\bf mainly because of their extremely poor knowledge}. {\bf To do that a suitable trained network has to be built with the chosen
intrinsic dust properties, and the same input neurons described above. This is far more convenient than including also these properties as neurons, because of their large number, and also because of little use for semi-analytical galaxy models whose range of predictions do not reach such details. This would serve to test specific requirements.}

\subsection{Computing performance}

The implementation of the ANN into \texttt{GRASIL} dramatically reduces
the CPU time required to run the code. As recalled above, with a $\sim 2$ GHz CPU
a single run could take anywhere up to $\sim 10$ minutes to calculate a SED, depending on the geometry.
With the use of an ANN this time is reduced to just a few seconds, with the main CPU time taken up by the
processes which are not calculated by the ANN; a CPU gain of more than 2
orders of magnitude. Such a remarkable reduction in computing costs should
make possible an efficient comparison of the SEDs of SAMs to large observational galaxy surveys
with a proper dust treatment.


\section{Applications}
\label{section:results}

\begin{figure}
  \centerline{\includegraphics[width=9cm]{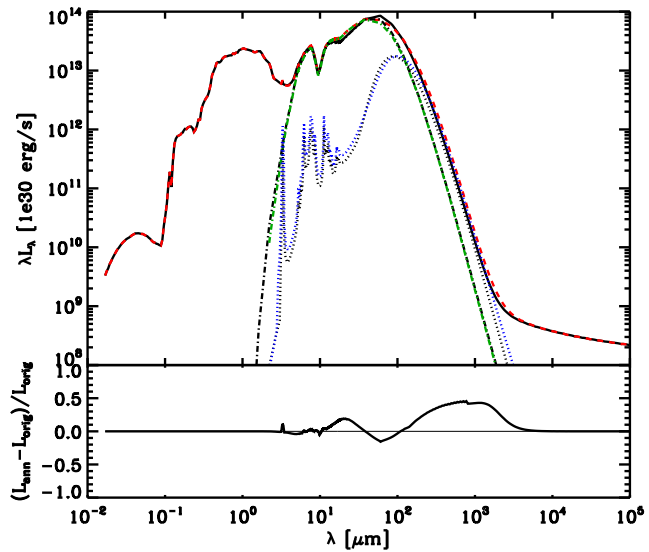}}
  \vspace{-3.3cm}
  \caption{M82: Original vs ANN and residuals. Upper panel: Black continuous is
  total original, red dashed is total ANN; dot-dashed black and green is for
  molecular clouds; dotted black and blue is cirrus. Lower panel: total residual}
\label{fig:cfrm82}
\end{figure}

\begin{figure}
  \centerline{\includegraphics[width=9cm]{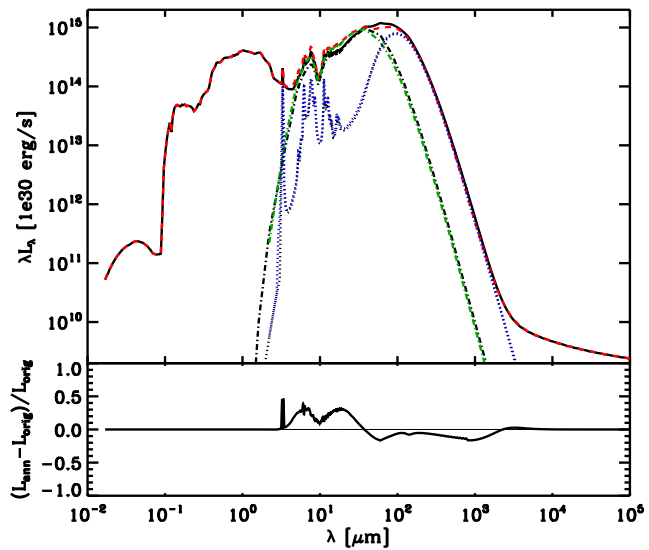}}
\vspace{-3.3cm}
  \caption{NGC6090: Original vs ANN and residual}
\label{fig:cfrn6090}
\end{figure}

\begin{figure}
  \centerline{\includegraphics[width=9cm]{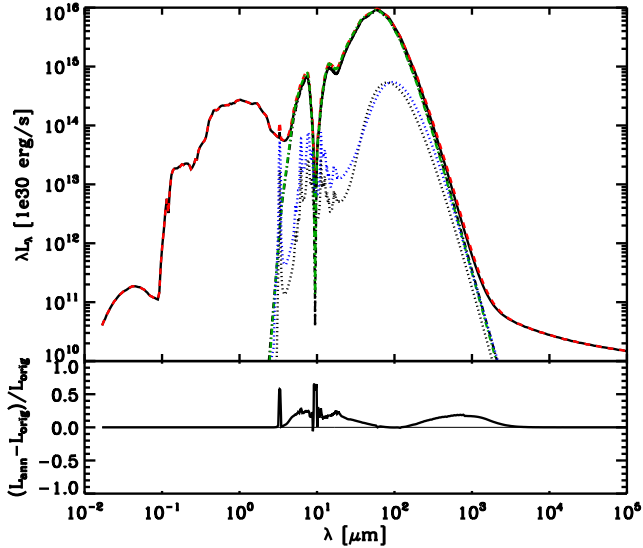}}
\vspace{-3.3cm}
  \caption{ARP220: Original vs ANN and residual}
\label{fig:cfrarp}
\end{figure}

\begin{figure}
  \centerline{\includegraphics[width=9cm]{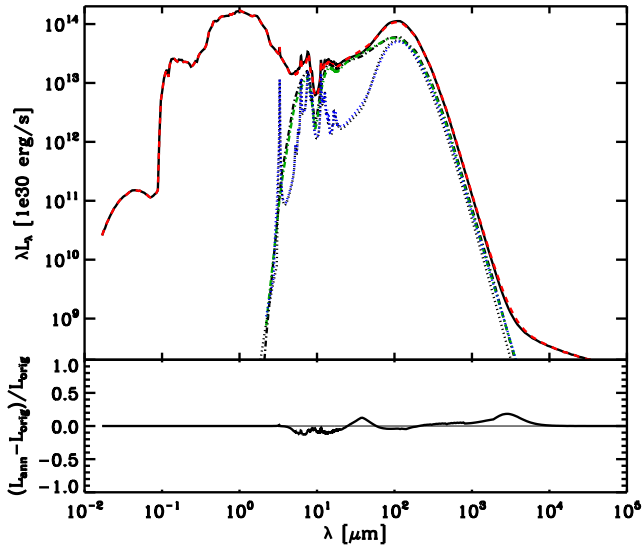}}
\vspace{-3.3cm}
  \caption{M51: Original vs ANN and residuals}
\label{fig:cfrm51}
\end{figure}

\begin{figure}
  \centerline{\includegraphics[width=9cm]{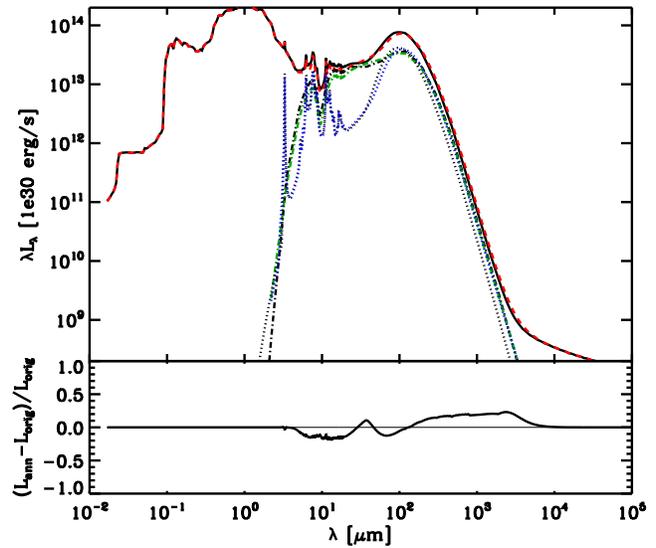}}
\vspace{-3.3cm}
  \caption{M100: Original vs ANN and residual}
\label{fig:cfrm100}
\end{figure}

\begin{figure}
  \centerline{\includegraphics[width=9cm]{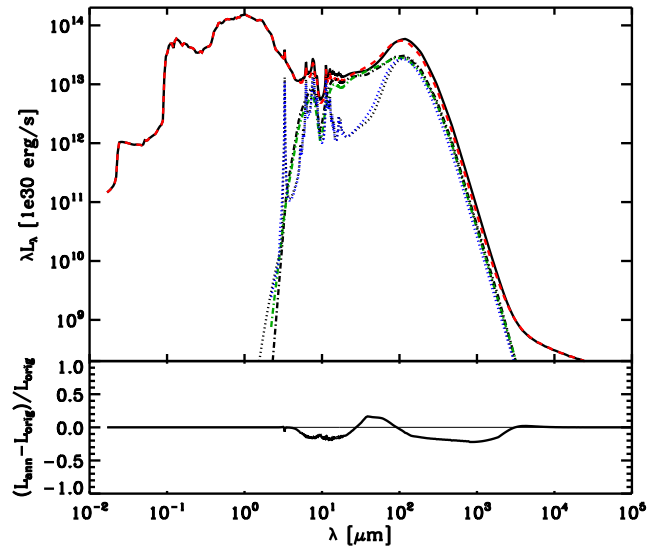}}
\vspace{-3.3cm}
  \caption{NGC6946: Original vs ANN and residual}
\label{fig:cfrn6946}
\end{figure}

\subsection{Examples with single SEDs}
\label{section:seds}

In Figs. \ref{fig:cfrm82} to \ref{fig:cfrn6946} we show examples of
comparisons between the SEDs directly computed with \texttt{GRASIL}
with those estimated with the ANN. These examples comprise model fits
to the real and well defined SEDs of galaxies in different
evolutionary states, which are commonly used as benchmarks for models
of dusty galaxies. The specific set of parameters of these models were
not included in the training set. Therefore the trained MLPs perform
well enough for most purposes, such as fast exploration of parameter
space. Given the small amount of CPU time to calculate a single SED
using the ANN, it will be possible to employ techniques of automatic
optimization for model parameters with suitable programs (e.g. MRQMIN
in Press et al.\ 1996).

\subsection{Application to the ABC semi-analytical model}
\label{section:abc}

As already mentioned in the Introduction, perhaps the most obvious
application in which a significantly quicker way to estimate a
\texttt{GRASIL} SED is a considerable advantage is when this model is
used in combination with SAMs. In this case, to test these models
against observations such as luminosity functions, number counts and
galaxy scaling relations, \texttt{GRASIL} has to be run for at least a
few thousand mock galaxies, a quite demanding computing task. On the
other hand, since many of these observables are integrated quantities,
reasonably small inaccuracies in the computation of each single SED,
without systematics, are acceptable.  In this Section we demonstrate
that our trained MLPs meet this practical request, showing some
applications with the \texttt{ABC} model (Granato et al.\ 2004) for
the co-evolution of spheroids and QSO, requiring only a spherical
geometry.  \texttt{ABC} is a simple, yet quite successful,
semi-analytical model originally developed to provide an
interpretation for sub-mm selected galaxies and their possible
descendants, the local massive spheroidal galaxies, accounting in
particular for the growth by accretion of a central super massive
black hole and its feedback on the host galaxy. The general behavior
of the evolution of the mock galaxies envisaged by the model is
characterized by a strong and relatively short dust-enshrouded SF phase
during which a central SMBH grows, a QSO phase halting subsequent star
formation, and then essentially passive evolution. We refer to the
original papers for more details (G04; Granato et al. 2006; Silva et
al. 2005; Lapi et al. 2006).

\begin{figure*}
  \centerline{\includegraphics[width=18cm]{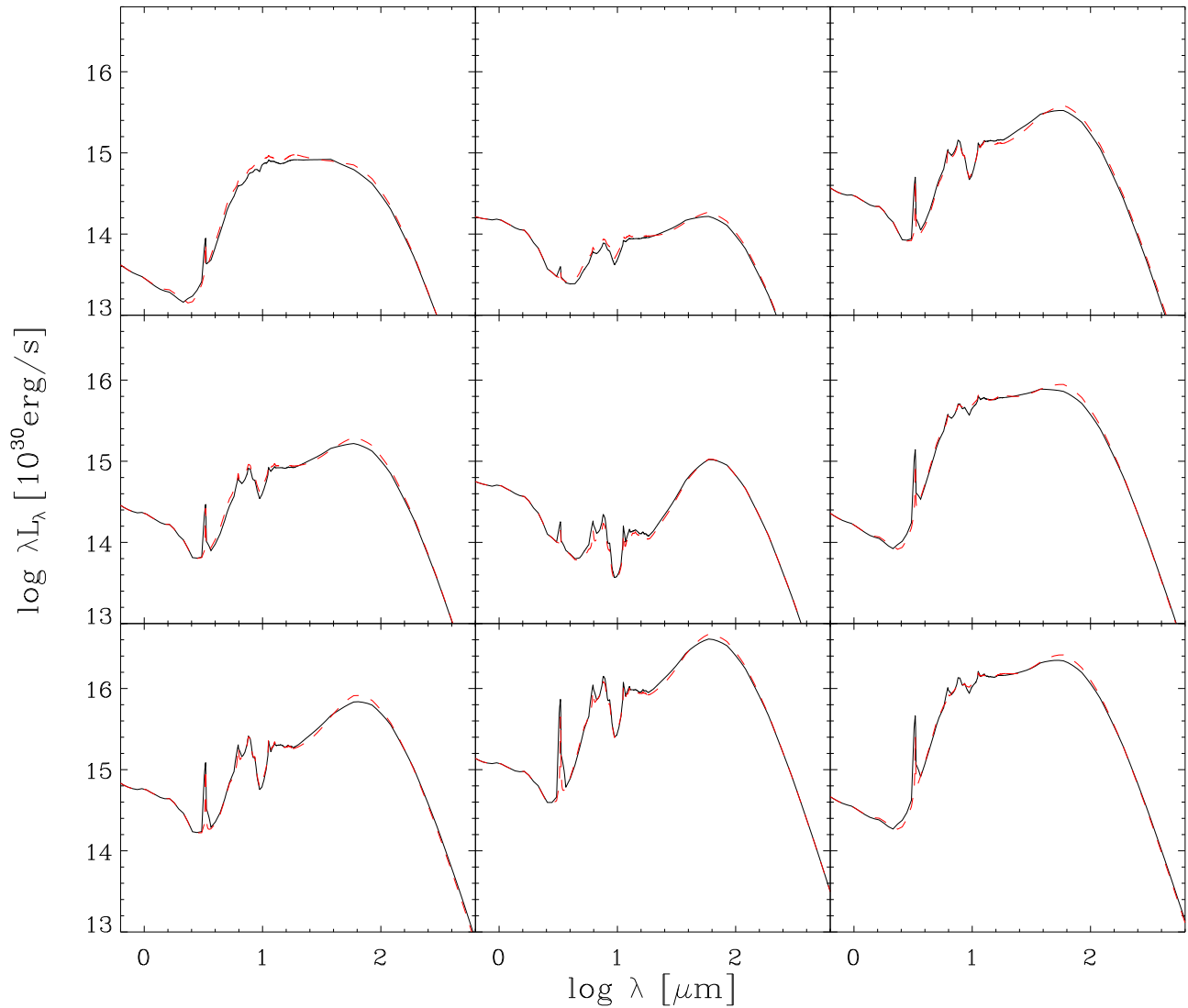}}
  \caption{Examples of randomly extracted galaxy models from the \texttt{ABC} SAM. The
  SEDs obtained either with the full \texttt{GRASIL} (dashed lines) or the ANN reconstruction (solid lines) are almost
  superimposed.}
  \label{fig:abcann}
\end{figure*}

\begin{figure}
  \centerline{\includegraphics[width=10cm,height=9cm]{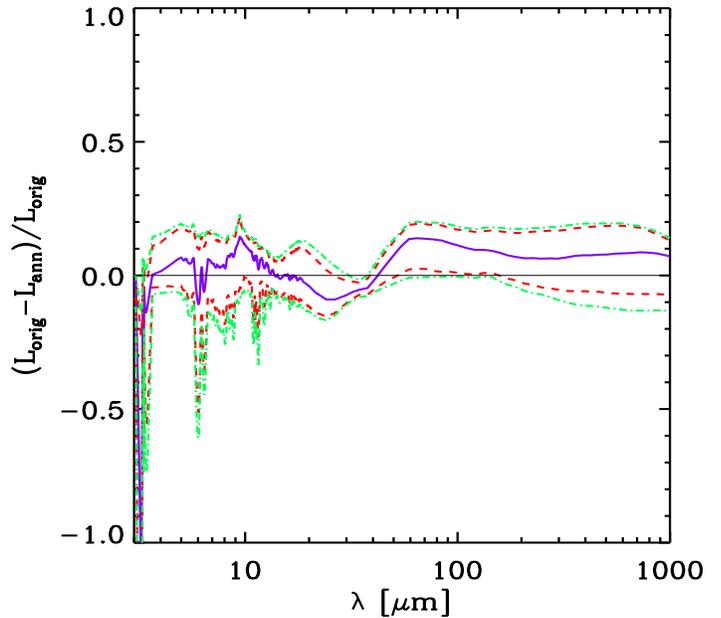}}
\caption{Residuals as a function of wavelength between original and ANN SEDs,
for a sample of about 400 dusty mock galaxies at various redshifts between 2 and 6,
generated by the \texttt{ABC} spheroid-QSO co-evolution model (Granato et al.\ 2004). Continuous (purple)
line: median; Dashed (red) lines: 10-90\% percentiles; Dot-dashed (green)
lines: 5-95\% percentiles.}
\label{fig:abcres}
\end{figure}

Here, to exemplify the effectiveness of our ANN computed SEDs in the
process of SAM validation, we show the expected number density of the
proto-spheroids output by \texttt{ABC} in the PACS and SPIRE {\it
Herschel} imaging bands, by computing the SED of each mock galaxy at
all phases (i.e. redshift slices), either with the full code or the ANN,
and compare the results. In Fig.\ \ref{fig:abcann} we show examples of
randomly extracted SEDs from the \texttt{ABC} model. The original and
ANN SEDs are often very difficult to distinguish in these plots.
A systematic comparison is in Fig.\ \ref{fig:abcres}, showing the residuals between the original and the ANN
SEDs vs wavelength for $\sim$ 400 objects extracted from \texttt{ABC}
galaxy catalogues at various redshifts between 2 and 6. We show the median and the $0.1-0.9$ and the $0.05-0.95$ percentiles. {\bf It is worth to point out that the typical error introduced by the use of the ANN is less than $\sim 10 \%$, meaning that it is likely dominated by uncertainties if the adopted \texttt{GRASIL} physics, by the simplified geometry and by the numerical approximations. However, a detailed comparison between these contributions to the total uncertainty in the model is outside the scope of this paper, while our point here is to investigate the capability of the ANN to avoid the time consuming \texttt{GRASIL} computations. }

\begin{figure}
\centerline{\includegraphics[width=10cm,height=7cm]{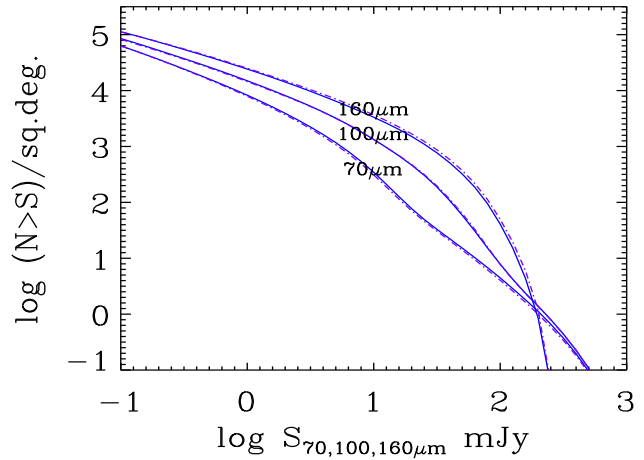}}
  \caption{Spheroidal integral galaxy number counts in the three {\it Herschel} PACS bands
  at $70$, $100$ and $160 \mu$m, as predicted by the \texttt{ABC} SAM (Granato et al.\ 2004): comparison
  between counts obtained with the full computation for the SED (dot-dashed violet line), and the ANN
  reconstruction (continuous blue line). The two lines are almost superimposed.}
  \label{fig:abcanncountpacs}
\end{figure}

\begin{figure}
\centerline{\includegraphics[width=10cm,height=7cm]{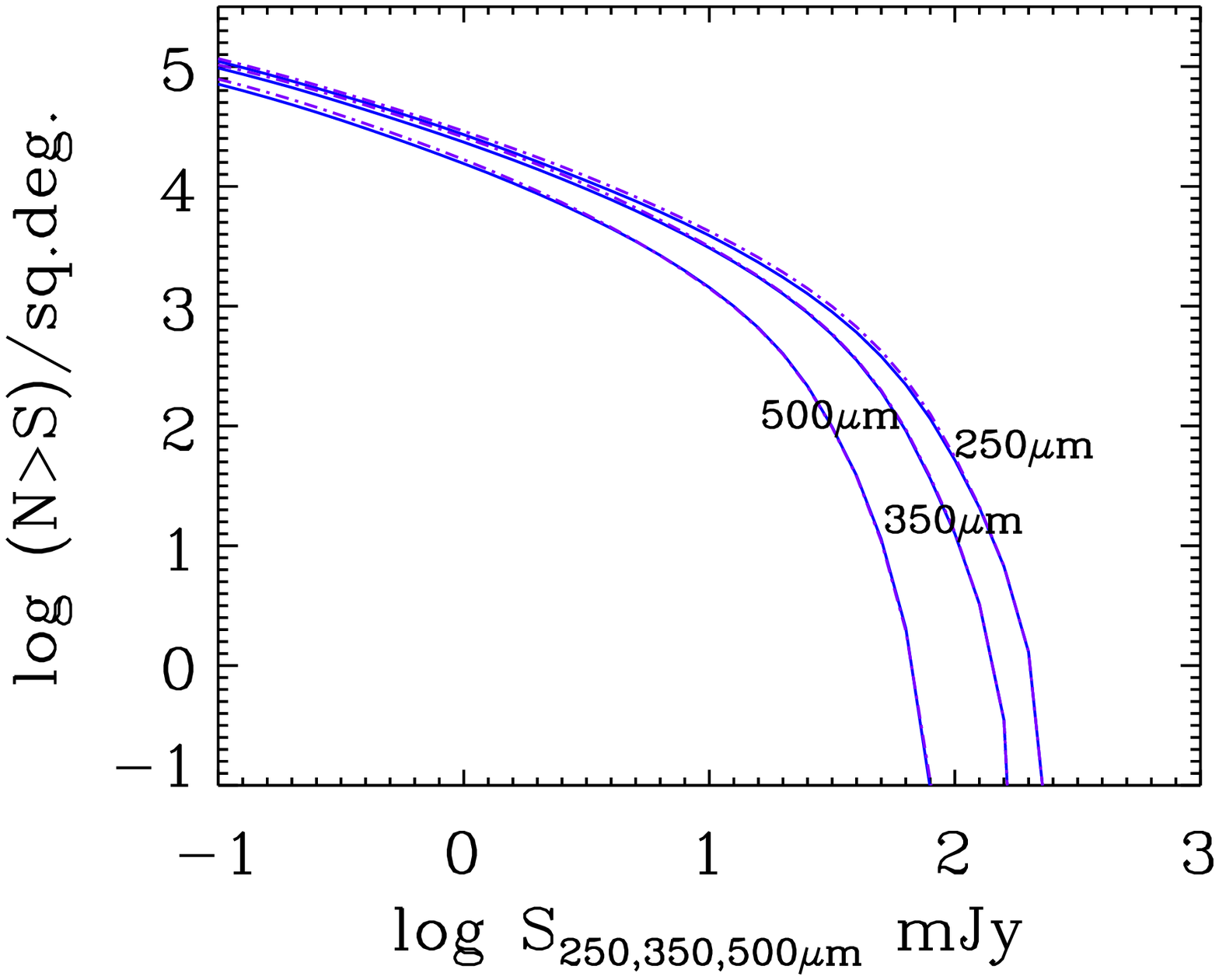}}
  \caption{Spheroidal integral galaxy number counts in the three {\it Herschel} SPIRE bands
  at $250$, $350$ and $500 \mu$m. Meaning of lines as in Fig.\ \ref{fig:abcanncountpacs}.}
  \label{fig:abcanncountspire}
\end{figure}

\begin{figure}
\centerline{\includegraphics[width=9.5cm,height=7cm]{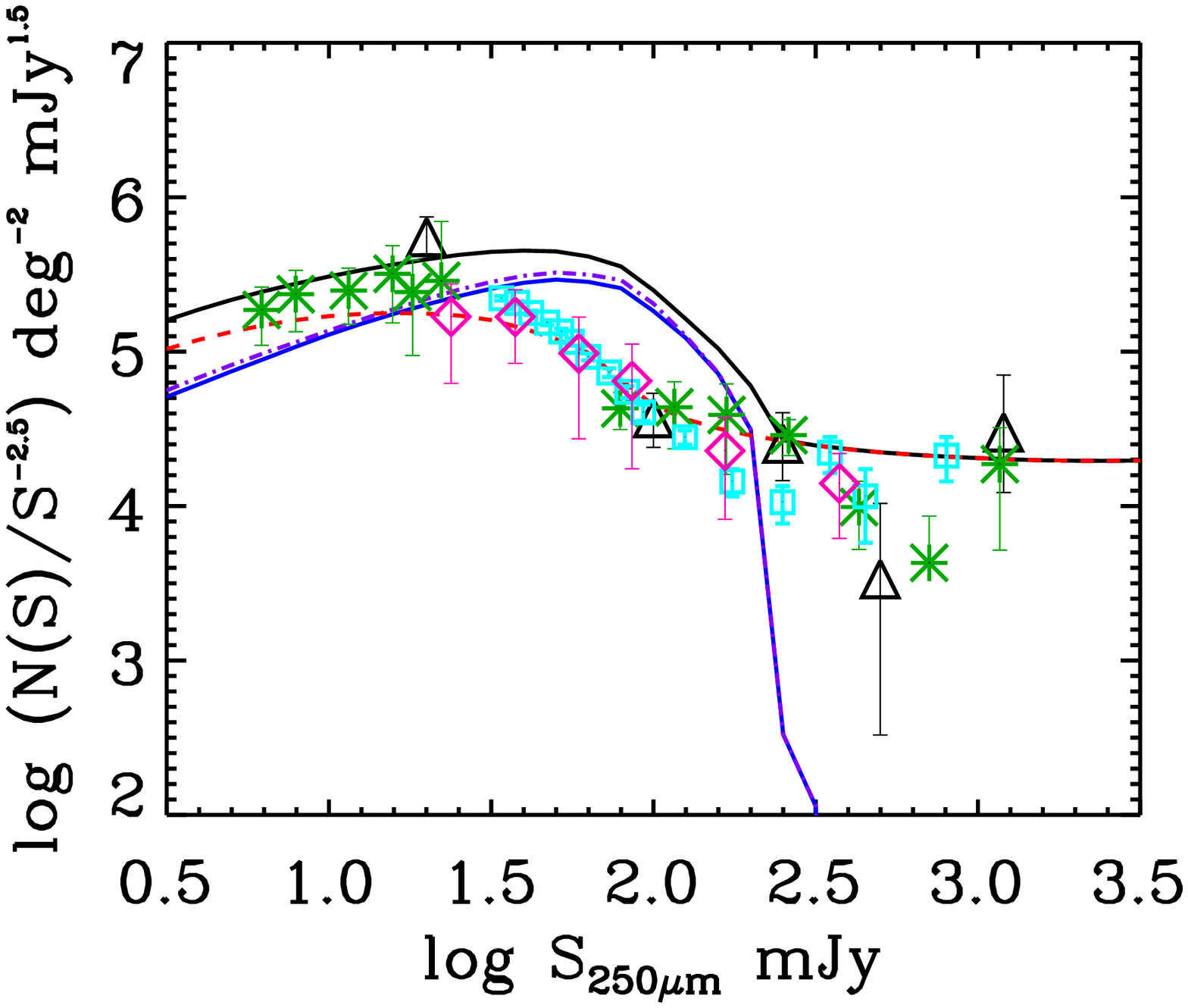}}
\centerline{\includegraphics[width=9.5cm,height=7cm]{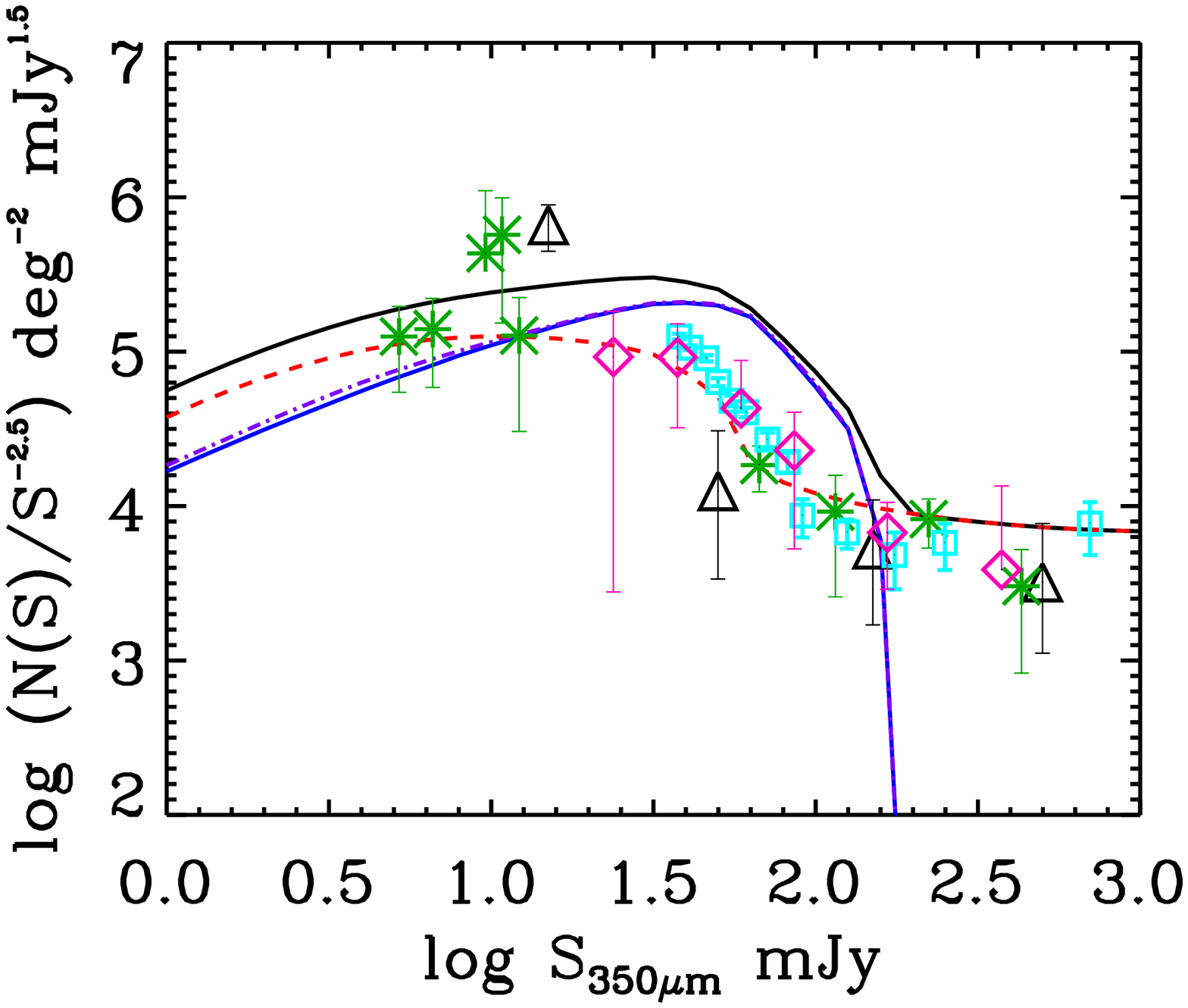}}
\centerline{\includegraphics[width=9.5cm,height=7cm]{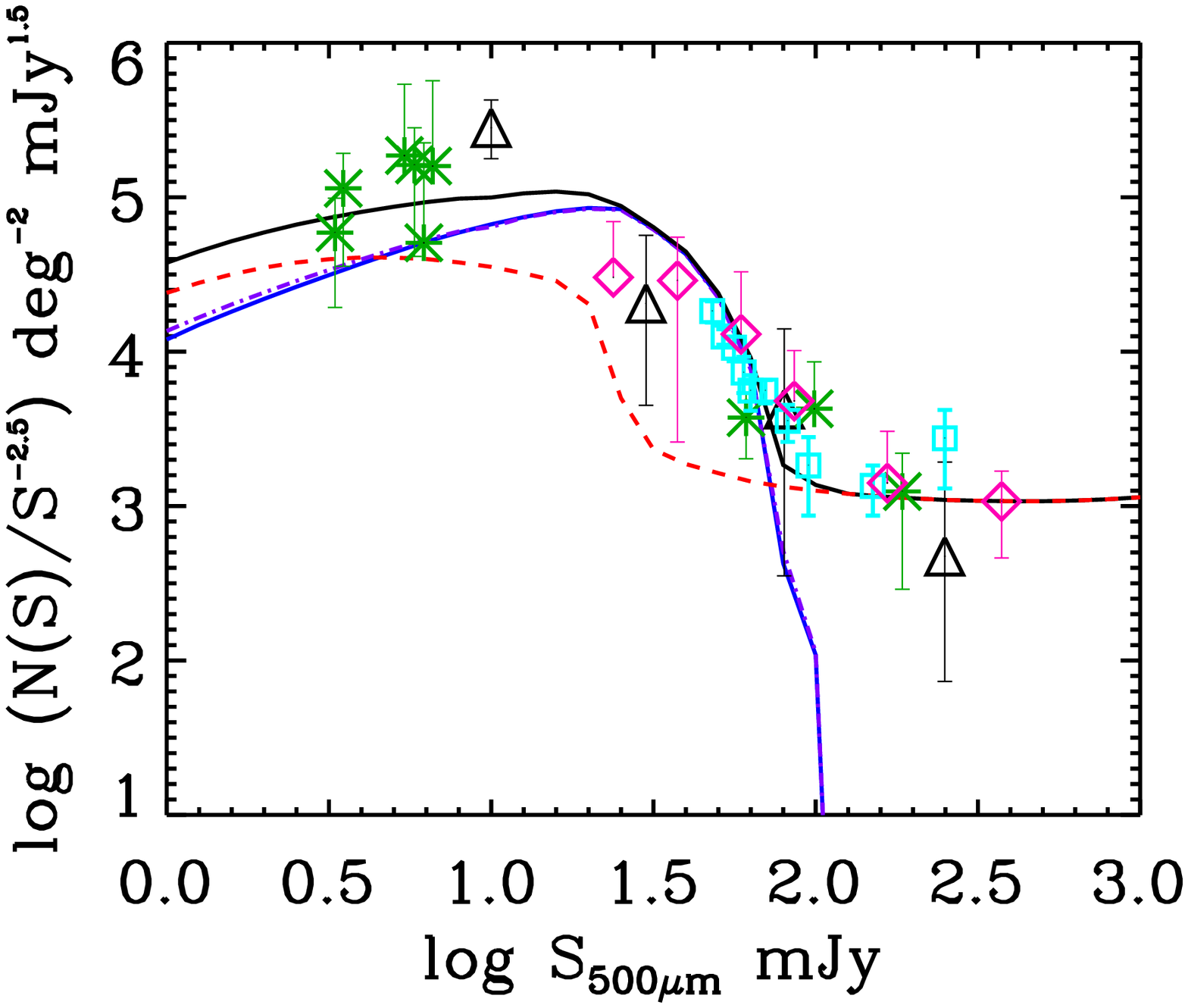}}
  \caption{Differential galaxy counts normalized to Euclidean
  at $250$, $350$ and $500 \mu$m. The 3 dot-dashed (violet) and continuous (blue) curves
  are the star forming spheroids from the G04 model with the full and ANN computation of the SEDs, respectively.
  The dashed (red) curve is the contribution from late type galaxies (spirals and starbursts).
  The total is the black continuous line. BLAST counts are by Patanchon et al.\ (2009, triangles) and Bethermin et al.\ (2010, asterisks). {\it Herschel}-SPIRE counts are by Clements et al.\ (2010, squares) and Oliver et al.\ (2010, diamonds).}
  \label{fig:abcannblast}
\end{figure}

Integrated quantities such as luminosity functions and number counts
are more accurately reproduced than single SEDs, since small
differences in the SEDs tend to be smoothed out. To illustrate this,
in Fig.\ \ref{fig:abcanncountpacs} and \ref{fig:abcanncountspire} we
show the integral galaxy counts in the PACS and SPIRE
{\it Herschel} bands at 70, 100, 160, 250, 350 and 500 $\mu$m for the
\texttt{ABC} model, obtained with both the full computed SEDs and with
the ANN quick estimate. The curves can hardly be distinguished, so
that the latter is fully adequate to compare model predictions with
available and forthcoming data.
Data for differential number counts
are available at $250$, $350$ and $500$ $\mu$m wavelengths, obtained
with the balloon-borne BLAST telescope (Devlin et al.\ 2009) and very recently with {\it Herschel}-SPIRE.
In Fig.\ \ref{fig:abcannblast} we compare the counts by the \texttt{ABC} model,
as obtained with the full and the ANN computation for the SEDs, and we
also compare with BLAST data by Patanchon et al.\
(2009, triangles) and Bethermin et al.\ (2010, asterisks), and with SPIRE data by Clements et al.\ (2010, squares) and Oliver et al.\ (2010, diamonds). In addition to the forming spheroids, we have included an empirical estimate for
the contribution by late-type galaxies (spiral and starbursts) from Silva et al.\ (2004). The number density of proto-spheroids appears consistent, although quite high particularly
at 250 and 350 $\mu$m compared with the available data. A deeper investigation
would require to test the effects of different dust properties and
star-dust distributions on the predicted counts, and the
implementation of a fast algorithm for the SEDs allows to easily
perform this task, moreover by taking into account the effects on the
full wavelength range. A discussion on the interpretation of galaxy
counts is beyond the scope of this paper, and will be presented in
another paper with a more general SAM.

\section{Conclusions}
\label{section:concl}

We have presented the implementation of an artificial neural network algorithm
to compute the SEDs with the \texttt{GRASIL} code. The main aim is to have a reliable
radiative transfer computation for the theoretical SEDs and a short computing time, sufficiently short to be applied
to cosmological volumes populated by semi-analytical galaxy formation models. But of course this opens the possibility
of fast exploration of parameters to fit data. The main points of the paper are listed in the following:
\begin{itemize}

\item SEDs are complex and non linear functions of many galaxy
properties resulting from their star formation and assembly histories,
such as the age and metallicity distribution of the stars, the amount and
composition of gas and dust, the relative distribution of dust and stars
etc. A radiative transfer computation of the stellar radiation field
through the dust distribution to get the extincted stellar and dust emission
spectrum is a time-consuming task. The required time becomes
prohibitive in particular for applications involving simulations of
cosmological galaxy catalogues with semi-analytical galaxy formation
models, requiring thousands of mock galaxies at each redshift
slice. On the other hand, in order to exploit as much as possible all
available data to constrain models, it would preferable to maintain the
possibility to assign to each galaxy a SED that as much as possible
reflects its effective properties instead of relying on pre-defined
templates that may have nothing to do with the galaxy configuration.

\item ANN are tools particularly suited to approximate complex
non-linear functions. We have implemented a standard feed-forward
back-propagation ANN into the \texttt{GRASIL} model. The main
characteristics of this model were defined by the requirements of
having a relatively realistic representation of galaxies (in
particular by accounting for a two-phase dusty medium heated by stars
of different ages and for the temperature distribution of the dust) and an
acceptable (for many applications) computing time. The real bottleneck
to get the SED is the computation of the dust emission spectrum, since
it requires the computation of the distribution of the radiation field
at each point in the galaxy, and the ensuing dust temperature for each
type of grain. Therefore we have implemented the ANN to compute the
dust emission spectra, and separately for the star forming molecular
clouds and the diffuse medium due to their different properties. The
gain in computing time is more than 2 orders of magnitude.

\item To implement the ANN we have (a) identified the quantities
(input neurons) that {\em effectively} control the shape of the dust
emission SED from the two dusty components, and (b) trained the
network with a large set of pre-computed models covering a large range
of values of the input neurons. The input neurons are two for MCs
(optical depth and ratio of outer to sublimation radius), and 7 or 9
for cirrus emission, for spherical or disc geometry respectively
(optical depths, hardness of the radiation field, ratios of star to dust scale radii,
mass and bolometric luminosity of the diffuse dust). The network is meant to be of
general use, because by construction it is independent of the
specifics of the galaxy formation model in use, the quantities that
effectively determine the shape of the dust emission spectrum are
extracted from the input star formation histories and used as input
neurons. This is complementary to the work by Almeida et al.\ (2010),
where an ANN has been implemented specifically for the combined
\texttt{GALFORM}+\texttt{GRASIL} model. In this case the ANN is meant
to give the full SED and the method is yet faster than the one
presented here.

\item We have tested the computation of the SEDs with the ANN with
single SEDs and with a simple semi-analytical model. In this first
paper the ANN has been implemented for pure spherical or pure
disc geometries. The mixed bulge+disc geometry will be presented and
applied in another paper. We have compared the full and the ANN
computation for model SEDs that fit nearby well-observed starburst and
disc galaxies. We have then made the same comparison
for SEDs and galaxy counts for the \texttt{ABC} semi-analytical model by Granato et
al.\ (2004) for the joint formation of spheroids and QSOs. The ANN
appears to perform well in all the explored cases, which cover star
formation histories ranging from relatively quiescent spirals, to
extreme dust-enshrouded starbursts. It is also to be noted that small
inaccuracies in the SEDs are smoothed out when computing integrated
quantities such as LFs and galaxy counts. As for the latter, we have
shown that the counts in the PACS and SPIRE {\it Herschel} imaging
bands for the \texttt{ABC} model, as obtained with the full and ANN
computation are almost superimposed. This means that a thorough
exploration of the effects of different assumptions on the dust
properties, not output by the galaxy formation model but that must be
assumed for the SEDs, can easily be performed. A discussion on the
implication of these counts for the {\it Herschel} surveys is beyond
the scope of this paper, and will be discussed elsewhere. The
computation of SEDs with the ANN method appears robust and
computationally advantageous to analyze and test galaxy formation
models in cosmological volumes.
\end{itemize}


{}


\begin{thebibliography}{}

\bibitem[]{} Almeida, C., Baugh, C. M., Lacey, C. G., Frenk, C. S., Granato, G. L., Silva, L., Bressan, A., 2010, MNRAS, 402, 544
\bibitem[]{} Baes, M., Davies, J.I., Dejonghe, H., Sabatini, S., Roberts, S., Evans, R., Linder, S.M.; Smith, R. M., de Blok, W. J. G., 2003, MNRAS, 343, 1081
\bibitem[]{durha} Baugh C.M., Lacey C.G., Frenk C.S., Granato G.L., Silva L., Bressan A., Benson A.J., Cole S., 2005, MNRAS, 356 ,1191
\bibitem[]{} B{\'e}thermin, M., Dole, H., Cousin, M., Bavouzet, N., 2010, A\&A, 516, 43
\bibitem[]{} Bianchi, S., 2007, A\&A, 471, 765
\bibitem[]{} Bianchi, S., 2008, A\&A, 490, 461
\bibitem[]{} Bianchi, S., Ferrara, A., Giovanardi, C., 1996, ApJ, 465, 127
\bibitem[]{} Bishop C.M., 1995, Neural Networks for Pattern Recognition, Oxford University Press
\bibitem[]{} Blaizot, J., Guiderdoni, B., Devriendt, J. E. G., Bouchet, F. R., Hatton, S. J., Stoehr, F., 2004, MNRAS, 352, 571
\bibitem[]{} Bressan, A., Granato, G. L., Silva, L., 1998, A\&A, 332, 135
\bibitem[]{gs2} Bressan A., Silva L., Granato G.L., 2002, A\&A, 392, 377
\bibitem[]{} Bundy, K., Ellis, R.S., Conselice, C.J., 2005, ApJ, 625, 621
\bibitem[]{} Caputi, K. I., McLure, R. J., Dunlop, J. S., Cirasuolo, M., Schael, A. M., 2006, MNRAS, 366, 609
\bibitem[]{} Chakrabarti, S., Fenner, Y., Cox, T. J., Hernquist, L., Whitney, B.A., 2008, ApJ, 688, 972
\bibitem[]{chael} Chary R., Elbaz D., 2001, ApJ, 556, 562
\bibitem[]{} Cimatti, A., Daddi, E., Renzini, A., Cassata, P., Vanzella, E., Pozzetti, L., Cristiani, S., Fontana, A., et al., 2004, Nature, 430, 184
\bibitem[]{} Cimatti, A., Daddi, E., Renzini, A., 2006, A\&A, 453, 29
\bibitem[]{} Clements, D. L. et al., 2010, A\&A, accepted, (arXiv1005.2409)
\bibitem[]{} Cole, S., Lacey, C. G., Baugh, C. M., Frenk, C. S., 2000, MNRAS, 319, 168
\bibitem[]{} Condon, J. J., 1992, ARA\&A
\bibitem[]{} da Cunha, E., Charlot, S., Elbaz, D., 2008, MNRAS, 388, 1595
\bibitem[]{danny} Dale, D.A., Helou, G., Contursi, A., Silbermann, N.A., Kolhatkar, S., 2001, ApJ, 549, 215
\bibitem[]{danny} Dale D.A., Helou G., 2002, ApJ, 576, 159
\bibitem[]{} Devlin M. J. et al., 2009, Nat, 458, 737
\bibitem[]{dgs99} Devriendt J.E.G., Guiderdoni B., Sadat, R., 1999, A\&A, 350, 381
\bibitem[]{} Dopita, M.A., Groves, B.A., Fischera, J., Sutherland, R.S., Tuffs, R.J., Popescu, C.C., Kewley, L.J., Reuland, M., Leitherer, C., 2005, ApJ, 619, 755
\bibitem[]{} Dopita, M.A., Fischera, J., Sutherland, R.S., Kewley, L.J., Tuffs, R.J., Popescu, C.C., van Breugel, W., Groves, B.A., Leitherer, C., 2006a, ApJ, 647, 244
\bibitem[]{} Dopita, M.A., Fischera, J., Sutherland, R.S., Kewley, L.J., Leitherer, C., Tuffs, R.J., Popescu, C.C., van Breugel, W., Groves, B.A., 2006b, ApJS, 167, 177
\bibitem[]{} Dorschner, J., Henning, T., 1995, A\&ARv, 6, 271
\bibitem[]{} Draine, B. T., 2003, ARA\&A, 41, 241
\bibitem[]{} Draine, B. T., Lee, H. M., 1984, ApJ, 285, 89
\bibitem[]{} Draine, B. T., Li, A., 2007, ApJ, 657, 810
\bibitem[]{} Drory, N., Bender, R., Feulner, G., Hopp, U., Maraston, C., Snigula, J., Hill, G. J., 2003, ApJ, 595, 698
\bibitem[]{} Drory, N., Salvato, M., Gabasch, A., Bender, R., Hopp, U., Feulner, G., Pannella, M., 2005, ApJ, 619, 131
\bibitem[]{} Efstathiou, A., Rowan-Robinson, M., Siebenmorgen, R., 2000, MNRAS, 313, 734
\bibitem[]{} Efstathiou, A., Rowan-Robinson, M., 2003, MNRAS, 343, 322
\bibitem[]{} Fabbiano, G., 2006, ARA\&A, 44, 323
\bibitem[]{} Fixsen, D. J., Dwek, E., Mather, J. C., Bennett, C. L., Shafer, R. A., 1998, ApJ, 508, 123
\bibitem[]{} Fontana, A., Pozzetti, L., Donnarumma, I., Renzini, A., Cimatti, A., Zamorani, G., Menci, N., Daddi, E., et al., 2004, A\&A, 424, 23
\bibitem[]{} Fontanot, F., Monaco, P., Silva, L., Grazian, A., 2007, MNRAS, 382, 903
\bibitem[]{} Fontanot, F., Somerville, R.S., Silva, L., Monaco, P., Skibba, R., 2009, MNRAS, 392, 553
\bibitem[]{} Fontanot, F., Monaco, P., 2010, MNRAS, accepted, arXiv:0911.5346
\bibitem[]{} Galliano, F., Madden, S. C., Jones, A. P., Wilson, C. D., Bernard, J.-P., Le Peintre, F., 2003, A\&A, 407, 159
\bibitem[]{gs2} Galliano, E., Alloin, D., Pantin, E., Granato, G. L., Delva, P., Silva, L., Lagage, P. O., Panuzzo, P., 2008, A\&A, 492, 3
\bibitem[]{} Genzel, R., Cesarsky, C.J., 2000, ARA\&A, 38, 761	
\bibitem[]{} Giavalisco, M., 2002, ARA\&A, 40, 579
\bibitem[]{} Gordon, K.D., Misselt, K. A., Witt, A. N., Clayton, G. C., 2001, ApJ, 551, 269
\bibitem[]{} Granato, G. L., Danese, L., 1994, MNRAS, 268, 235
\bibitem[]{} Granato, G. L., Danese, L., Franceschini, A., 1997, ApJ, 486, 147
\bibitem[]{} Granato, G. L., Lacey, C. G., Silva, L., Bressan, A., Baugh, C. M., Cole, S., Frenk, C. S., 2000, ApJ, 542, 710 [G00]	
\bibitem[]{} Granato, G. L., De Zotti, G., Silva, L., Bressan, A., Danese, L., 2004, ApJ, 600, 580 [G04]
\bibitem[]{} Granato, G. L., Silva, L., Lapi, A., Shankar, F., De Zotti, G., Danese, L., 2006, MNRAS, 368L, 72
\bibitem[]{} Groves, B., Dopita, M.A., Sutherland, R.S., Kewley, L.J., Fischera, J., Leitherer, C., Brandl, B., van Breugel, W., 2008, ApJS, 176, 438
\bibitem[]{} Guiderdoni, B., Hivon, E., Bouchet, F. R., Maffei, B., 1998, MNRAS, 295, 877
\bibitem[]{} Hatton, S., Devriendt, J. E. G., Ninin, S., Bouchet, F. R., Guiderdoni, B., Vibert, D., 2003, MNRAS, 343, 75
\bibitem[]{} Hauser, M.G., Dwek, E., 2001, ARA\&A, 39, 249
\bibitem[]{} Haykin, S., 1999, Neural Networks: A Comprehensive Foundation, 2nd ed.
(Englewood Cliffs, NJ: Prentice-Hall)
\bibitem[]{} Holland, W. S., Robson, E. I., Gear, W. K., Cunningham, C. R., Lightfoot, J. F., Jenness, T., Ivison, R. J., Stevens, J. A., et al., 1999, MNRAS, 303, 659
\bibitem[]{} Hughes, D.H., Serjeant, S., Dunlop, J., Rowan-Robinson, M., Blain, A., Mann, R.G., Ivison, R., Peacock, J., 1998, Nature, 394, 241
\bibitem[]{} Iglesias-Páramo, J., Buat, V., Hernández-Fernández, J., Xu, C. K., Burgarella, D., Takeuchi, T. T., Boselli, A., Shupe, D., et al., 2007, ApJ, 670, 279
\bibitem[]{} Jonsson, P., 2006, MNRAS, 372, 2
\bibitem[]{} Jonsson, P., Primack, J.R., 2010, NewA, 15, 509
\bibitem[]{} Jonsson, P., Groves, B.A., Cox, T.J., 2010, MNRAS, 403, 17
\bibitem[]{} Kang, X., Jing, Y. P., Mo, H. J., Börner, G., 2005, ApJ, 631, 21
\bibitem[]{} Kauffmann, G., Colberg, J. M., Diaferio, A., White, S. D. M., 1999, MNRAS, 303, 188
\bibitem[]{} Kessler, M. F., Steinz, J. A., Anderegg, M. E., Clavel, J., Drechsel, G., Estaria, P., Faelker, J., Riedinger, J. R., et al., 1996, A\&A, 315L, 27
\bibitem[]{} Kitzbichler, M. G., White, S. D. M., 2007, MNRAS, 376, 2	
\bibitem[]{} Lacey, C., Silk, J., 1991, ApJ, 381, 14
\bibitem[]{} Lacey, C., Cole, S., 1993, MNRAS, 262, 627	
\bibitem[]{} Lacey, C. G., Baugh, C. M., Frenk, C. S., Silva, L., Granato, G. L., Bressan, A., 2008, MNRAS, 385, 1155
\bibitem[]{} Lacey, C. G., Baugh, C. M., Frenk, C. S., Benson, A. J., Orsi, A., Silva, L., Granato, G. L.,
Bressan, A., 2010, MNRAS, 405, 2
\bibitem[]{} Lagache, G.; Dole, H.; Puget, J.-L., 2003, MNRAS, 338, 555	
\bibitem[]{} Lagache, G., Puget, J.-L., Dole, H., 2005, ARA\&A, 43, 727
\bibitem[]{} Laor, A., Draine, B. T., 1993, ApJ, 402, 441
\bibitem[]{} Lapi, A., Shankar, F., Mao, J., Granato, G. L., Silva, L., De Zotti, G., Danese, L., 2006, ApJ, 650, 42
\bibitem[]{} Li, A., Draine, B. T., 2001, ApJ, 554, 778
\bibitem[]{} Li, Y., Hopkins, P.F., Hernquist, L., Finkbeiner, D.P., Cox, T.J., Springel, V., Jiang, L., Fan, X., Yoshida, N., 2008, ApJ, 678, 41
\bibitem[]{} Maraston, C., 2005, MNRAS, 362, 799
\bibitem[]{} Mathis, J. S., 1990, ARA\&A, 28, 37
\bibitem[]{} McCulloch W., Pitts W., 1943, Bull. Math. Biophys., 5, 115
\bibitem[]{} Meurer, G.R., Heckman, T.M., Calzetti, D., 1999, ApJ, 521, 64
\bibitem[]{} Michalowski, M.J., Hjorth, J., Malesani, D., Michalowski, T., Castro Cerón, J.M., Reinfrank, R.F., Garrett, M.A., Fynbo, J. P. U., Watson, D. J., Jørgensen, U. G., 2009, ApJ, 693, 347
\bibitem[]{} Michalowski, M.J., Watson, D., Hjorth, J., 2010, ApJ, 712, 942
\bibitem[]{} Misiriotis, A., Popescu, C. C., Tuffs, R., Kylafis, N. D., 2001, A\&A, 372, 775
\bibitem[]{} Misselt, K. A., Gordon, Karl D., Clayton, Geoffrey C., Wolff, M. J, 2001, ApJ, 551, 277
\bibitem[]{} Möllenhoff, C., Popescu, C. C., Tuffs, R. J., 2006, A\&A, 456, 941
\bibitem[]{} Monaco, P., Fontanot, F., Taffoni, G., 2007, MNRAS, 375, 1189
\bibitem[]{} Narayanan, D., Hayward, C. C., Cox, T. J., Hernquist, L., Jonsson, P., Younger, J. D., Groves, B., 2010, MNRAS, 401, 1613
\bibitem[]{} Neugebauer, G., Habing, H. J., van Duinen, R., Aumann, H. H., Baud, B., Beichman, C. A., Beintema, D. A., Boggess, N., et al., 1984, ApJ, 278, 1
\bibitem[]{} Oliver, S. J. et al., 2010, A\&A, accepted, (arXiv1005.2184)
\bibitem[]{} Panuzzo, P., Bressan, A., Granato, G. L., Silva, L., Danese, L., 2003, A\&A, 409, 99
\bibitem[]{} Panuzzo, P., Granato, G. L., Buat, V., Inoue, A. K., Silva, L., Iglesias-Páramo, J., Bressan, A., 2007a, MNRAS, 375, 640
\bibitem[]{} Panuzzo, P., Vega, O., Bressan, A., Buson, L., Clemens, M., Rampazzo, R., Silva, L., Valdés, J. R., Granato, G. L., Danese, L., 2007b, ApJ, 656, 206
\bibitem[]{} Patanchon G. et al., 2009, ApJ, 707, 1750
\bibitem[]{} Popescu, C. C., Misiriotis, A., Kylafis, N. D., Tuffs, R. J., Fischera, J., 2000, A\&A, 362, 138
\bibitem[]{} Press, W.H., Teukolsky, S.A., Vetterling, W.T., Flannery, B.P., Michael Metcalf, M., 1996, Numerical Recipes in Fortran 90, Vol. 2,  Cambridge University Press
\bibitem[]{} Puget, J.-L., Abergel, A., Bernard, J.-P., Boulanger, F., Burton, W. B., Desert, F.-X., Hartmann, D., 1996, A\&A, 308, 5
\bibitem[]{} Rocha, M., Jonsson, P., Primack, J. R., Cox, T. J., 2008, MNRAS, 383, 1281
\bibitem[]{} Rojas R., 1996, Neural Networks - A Systematic Introduction, Springer-Verlag, Berlin
\bibitem[]{} Rosenblatt F., 1958, Psychol. Rev., 65, 386
\bibitem[]{} Rowan-Robinson, M., 1980, ApJS, 44, 403
\bibitem[]{} Rowan-Robinson, M., Crawford, J., 1989, MNRAS, 238, 523
\bibitem[]{} Rumelhart D. E., Hitton G. E., Williams R. J., 1986, Parallel Distributed
Processing. MIT Press, Cambridge, MA
\bibitem[]{} Santini, P., Maiolino, R., Magnelli, B., Silva, L., Grazian, A., Altieri, B., Andreani, P., Aussel H., et al., 2010, A\&AL, accepted (arXiv:1005.5678)
\bibitem[]{} Saracco, P., Longhetti, M., Severgnini, P., Della Ceca, R., Braito, V., Mannucci, F., Bender, R., Drory, N., et al., 2005, MNRAS, 357, 40
\bibitem[]{} Schurer, A., Calura, F., Silva, L., Pipino, A., Granato, G. L., Matteucci, F., Maiolino, R., 2009, MNRAS, 394, 2001
\bibitem[]{} Siebenmorgen, R.; Krügel, E., 2007, A\&A, 461, 445
\bibitem[]{sil98} Silva L., Granato G.L., Bressan A., Danese L., 1998, ApJ, 509, 103 [S98]
\bibitem[]{s2} Silva L., 1999, PhD thesis [S99]
\bibitem[]{s2} Silva, L., Granato, G. L., Bressan, A., Lacey, C.G., Baugh, C. M., Cole, S., Frenk, C. S., 2001, Ap\&SS, 276, 1073	
\bibitem[]{s2} Silva, L., De Zotti, G., Granato, G. L., Maiolino, R., Danese, L., 2004, arXiv:astro-ph/0403166
\bibitem[]{s2} Silva, L., De Zotti, G., Granato, G. L., Maiolino, R., Danese, L., 2005, MNRAS, 357, 1295
\bibitem[]{} Smail, I., Ivison, R. J., Blain, A. W., 1997, ApJ, 490L, 5
\bibitem[]{} Smail, I., Ivison, R. J., Blain, A. W., Kneib, J.-P., 2002, MNRAS, 331, 495
\bibitem[]{} Soifer, B. T., Neugebauer, G., Houck, J. R., 1987, ARA\&A, 25, 187
\bibitem[]{} Soifer, B. T., Helou, G., Werner, M., 2008, ARA\&A, 46, 201
\bibitem[]{} Somerville, R. S.; Primack, J. R., 1999, MNRAS, 310, 1087
\bibitem[]{} Springel, V., White, S. D. M., Jenkins, A., Frenk, C. S., Yoshida, N., Gao, L., Navarro, J., Thacker, R., et al. 2005, Nature, 435, 629
\bibitem[]{} Stasinska, G., 2007, arXiv:0704.0348
\bibitem[]{} Swinbank, A. M., Lacey, C. G., Smail, I., Baugh, C. M., Frenk, C. S., Blain, A. W., Chapman, S. C., Coppin, K. E. K., Ivison, R. J., Gonzalez, J. E., Hainline, L. J., 2008, MNRAS, 391, 420
\bibitem[]{} Takagi, T., Arimoto, N., Hanami, H., 2003, MNRAS, 340, 813
\bibitem[]{} Takagi, T., Vansevicius, V., Arimoto, N., 2003, PASJ, 55, 385
\bibitem[]{} Tuffs, R. J., Popescu, C. C., Völk, H. J., Kylafis, N. D., Dopita, M. A., 2004, A\&A, 419, 821
\bibitem[]{veg} Vega O., Silva L., Panuzzo P., Bressan A., Granato G.~L., Chavez M., 2005, MNRAS, 364, 1286
\bibitem[]{} Vega, O., Clemens, M. S., Bressan, A., Granato, G. L., Silva, L., Panuzzo, P., 2008, A\&A, 484, 631
\bibitem[]{} Werner, M. W., Roellig, T. L., Low, F. J., Rieke, G. H., Rieke, M., Hoffmann, W. F., Young, E., Houck, J. R., et al., 2004, ApJS, 154, 1
\bibitem[]{} White, S. D. M., Rees, M. J., 1978, MNRAS, 183, 341
\bibitem[]{} White, S. D. M.; Frenk, C. S., 1991, ApJ, 379, 52

\end{thebibliography}
\end{document}